\documentclass[pdftex,twocolumn,epjc3]{svjour3}          

\RequirePackage[T1]{fontenc}

\smartqed  

\RequirePackage{graphicx}
\RequirePackage{mathptmx}      
\RequirePackage{flushend}
\RequirePackage{bm}
\RequirePackage[numbers,sort&compress]{natbib}
\RequirePackage[colorlinks,citecolor=blue,urlcolor=blue,linkcolor=blue]{hyperref}

\journalname{Eur. Phys. J. C}

\begin{document}

\title{Exclusive semileptonic $B_c$-meson decays to radially excited charmonium and charm meson states.}

\author{Lopamudra Nayak\thanksref{e1,addr1}
	 \and	P. C. Dash\thanksref{addr1},
	 \and S. Kar\thanksref{e2,addr2}
	 \and N. Barik\thanksref{addr3} 
	}

\thankstext{e1}{lopalmn95@gmail.com}
\thankstext{e2}{skar09.sk@gmail.com}

\institute{ Department of Physics, Siksha $'O'$ Anusandhan (Deemed to be University), Bhubaneswar-751030, India\label{addr1}
          \and
          Department of Physics, Maharaja Sriram Chandra Bhanja Deo University, Baripada-757003, India\label{addr2}
          \and
          Department of Physics, Utkal University, Bhubaneswar-751004, India.\label{addr3}
}

\date{Received: date / Accepted: date}

\maketitle

\begin{abstract}
	In the wake of recent measurements of ratios of semileptonic branching fractions: ${\cal R}_{J/\psi}, {\cal R}_D$ and ${\cal R}_{D^*}$ reported by the LHCb, BELLE and HFLAV Collaborations, we calculate invariant form factors for the exclusive semileptonic $B_c$-meson decays to radially excited charmonium and charm meson states in the full kinematical region within the framework of relativistic independent quark (RIQ)model. We evaluate the lepton mass effect in the decay processes induced by $b\to c,u$ transition at the quark level. Our predictions on branching fractions for $B_c\to \eta_c/\psi(nS)$ are found $\sim10^{-2}-10^{-4}$ and that for $B_c\to D^*(nS)$ are $\sim 10^{-4}$ in their $e^-$ decay modes, which lie within the detection accuracy of current experiment. Our predictions on branching fractions, forward backward asymmetry and asymmetry parameter are found in reasonable agreement with other model predictions; which can hopefully be tested in future experiments at LHC and Tevatron. Our predicted observable ${\cal R}$ in this sector are found comparable to other standard model (SM) predictions that violate the lepton flavor universality hinting at new physics beyond SM. 
	
\end{abstract}

\section{Introduction}
One of the most engrossing puzzles in flavor physics in recent years has been the observed deviations of observables in the semileptonic $B_c$-decays from the corresponding SM predictions. The measurements of observables ${\cal R}_{J/\psi}$: the ratio of semileptonic branching fractions reported by the LHCb Collaboration\cite{A1}:
\begin{eqnarray}
	{\cal R}_{J/\psi}=\frac{{\cal B}(B_c\to J/\psi \tau \nu_{\tau})}{{\cal B}(B_c\to J/\psi \mu \nu_{\mu})}=0.71\pm0.17(stat)\pm 0.18(syst)\nonumber
\end{eqnarray}
lie within $2\sigma$ above the range of existing SM predictions \cite{A2,A3,A4,A5,A6}. A series of measurements of identical decay modes $B\to D^*l\nu_l$ by BaBar \cite{A7,A8}, BELLE \cite{A9,A10,A11}, and LHCb \cite{A12,A13} also mark significant deviation of observables from the SM predictions. The average of the observed ${\cal R}_D$ and ${\cal R}_{D^*}$ by the HFLAV Collaboration \cite{A14} which read

\begin{eqnarray}
	{\cal R}_D=\frac{{\cal B}(B\to D \tau \nu_{\tau})}{{\cal B}(B\to D \mu \nu_{\mu})}=0.407\pm0.39\pm0.024\nonumber\\
	{\cal R}_D=\frac{{\cal B}(B\to D^* \tau \nu_{\tau})}{{\cal B}(B\to D^* \mu \nu_{\mu})}=0.304\pm0.013\pm0.007\nonumber,
\end{eqnarray}

\noindent reveal deviations from the SM predictions of ${\cal R}_D\simeq 0.299\pm0.011$ from lattice calculation\cite{A15,A16} and ${\cal R}_{D^*}\simeq 0.252\pm0.003$ from model calculation \cite{A17}, respectively. Several experimental studies on $B_c$-decays are cited in the literature, most of which pertain to $B_c$ decays to final mesons in their ground state only. This includes the observation of $B_c$ decays reported by the CDF Collaboration \cite{A18,A19,A20} and the observation of the decay modes: $B_c\to J/\psi\mu\nu_{\mu}X$ \cite{A21}, where X denotes any possible additional particle in the final state, yielding more precise measurement of $B_c$ lifetime: $\tau_{B_c}=0.51^{+0.18}_{-0.16}(stat)\pm0.03(syst)ps$ and its mass: $M=6.40\pm0.39\pm0.13\ GeV$. The radially excited $\psi(2S)$ and $\eta_c(2S)$ states in the heavy flavored charmonium sector and many charmonium-like $D^0(2S)$ and $D^{*0}(2S)$ states have also been detected. \\

	Several theoretical studies on semileptonic $B_c$-decays to charmonium and charm meson states have been cited in the literature under different theoretical approaches. A few noteworthy among them are the potential model\cite{A22}, non-relativistic quark model (NRQM) \cite{A5}, the Baur-Stech-Wirbel (BSW) approach \cite{A23}, QCD sum rules (QCDSR) \cite{A3,A24,A25,A26,A27,A28}, non-relativistic QCD (NRQCD) \cite{A29,A30,A31,A32,A33}, studies based on the Bethe-Salpeter Equation\cite{A34,A35}, Bethe-Salpeter relativistic quark model \cite{A36}, and improved Bethe-Salpeter method \cite{A37}, relativistic quark model (RQM) \cite{A12,A38,A39,A40,A41}, relativistic constituent quark model (RQCM) on light front \cite{A42,A43}, Covariant confined quark model (CCQM)\cite{A44,A45,A46}, Covariant light-front quark model (CLQM)\cite{A47}, QCD potential model (QCDPM) \cite{A48,A49,A50}, perturbative QCD (pQCD) \cite{A51,A52,A53,A54,A55,A56,A57,A58,A59,A60,A61}, Lattice QCD \cite{A62}, Light-cone QCD sum rules \cite{A63,A64}, and the light-front quark model (LFQM) \cite{A65}. The detection of the semileptonic $B_c$-decay modes to radially excited charmonium and charm meson states, which are easier to identify in the experiment and the deviation of observables ${\cal R}$ from the SM predictions provide us necessary motivation to study these decay modes in the RIQ model framework.\\
	The RIQ model, developed by our group, has been applied extensively in the description of wide-ranging hadronic phenomena including the static properties of hadrons \cite{A66} and several decay properties such as the radiative; weak radiative, rare radiative; leptonic, weak leptonic, radiative leptonic; and non-leptonic decays. The semileptonic decays of heavy flavored mesons in their ground states have also been predicted within the framework of the RIQ model.\\
		It may be noted that our predictions of semileptonic decay modes \cite{A67,A68,A69,A70} are based on vanishing lepton mass limit, where only the 3-vector (or space component) hadronic current form factors contribute to the decay amplitude. The scalar-(or time component) hadronic current form factor is not accessible in such descriptions. However, in the description of the semitauonic decay modes involving nonvanishing lepton mass, one needs to consider both the space- and time component of hadronic current form factors that contribute to the decay amplitude. In our recent analysis of exclusive semileptonic $B_c$ meson decays to the charmonium and charm meson in their ground states, we predict the lepton mass effect by evaluating the $B_c$ decays to the electronic as well as tauonic modes\cite{A71}. Inspired by our recent prediction on exclusive nonleptonic $B_c$ meson decays to radially excited charmonium and charmonium like states\cite{A72}, we would like to extend the applicability of our model to analyse the semileptonic $B_c$ decays to the charmonium and charm mesons in their radially excited $(nS)$ states, where $n=2,3$ and evaluate the lepton mass effect in those decay processes. In the process we intend to predict the branching fractions, ratio of branching fraction ${\cal R}$ in comparison to those obtained in our previous analysis\cite{A71}. We ignore the decay channels involving higher $4S$ final states since their properties are still not understood well.\\
		
		The semileptonic decay modes:$B_c\to \eta_c(J/\psi)l\nu_l$ and $B_C\to D(D^*)l\nu_l$ are induced by $b\to cl\nu_l$ and $b\to ul\nu_l$ transitions at the quark level, respectively. The kinematic range of $q^2$ for $B_c\to \eta_c(J/\psi)l\nu_l$ is $0\leq q^2\le 10\ GeV^2$, and that for $B_c\to D(D^*)l\nu_l$ is $0\leq q^2\le 18\ GeV^2$. In the $B_c$-meson rest frame, the maximum recoil momenta of the final state charmonium ($\eta_c,\ J/\psi$) and charm ($D,\ D^*$) meson states are estimated to be in the same range of magnitude as their masses. With this kinematic constraint, it is interesting to analyze here the decay modes $B_c\to \eta_c(J/\psi)l\nu_l$ and $B_c \to D(D^*)l\nu_l$ separately. Due to their simple theoretical description via a tree-level diagram in the SM, the effects of the strong interaction can be separated from the effects of the weak interaction into a set of Lorentz-invariant form factors. The study of semileptonic decays, therefore, reduces to a calculation of relevant weak form factors in a suitable phenomenological model framework.\\
		
		  It is worthwhile to note here following few points: (1) In some of the theoretical approaches like the pQCD and QCD sum rules, for example, the form factors are determined first, with an end point normalization at minimum $q^2$ (maximum recoil) or maximum $q^2$ (minimum recoil) and then extrapolated to the entire kinematical region using some monopole/ dipole /Gaussian ansatz. This makes the form factor estimation less reliable. In the present analysis, the relevant form factors are evaluated in the full kinematical range of $q^2$, which makes the predictions of physical observables more accurate. (2) Since the decay modes considered here involve both the light and heavy leptons, we would evaluate both the vector- and scalar components of hadronic current form factors that contribute to the decay amplitude, study the lepton mass effects in such decays and predict the observables ${\cal R}$ in comparison with other SM predictions. (3) We update some input hadronic parameters according to the Particle Data Group \cite{A73} for our numerical analysis.\\
		The outline of the paper is as follows. In Sec. 2 we briefly discuss the general formalism and kinematics for semileptonic $B_c$-meson decays. Section 3 describes the extraction of the transition form factors from the invariant transition amplitudes in the framework of the RIQ model. Section 4 is devoted to the numerical analysis of the form factors, decay rates, branching fractions, ratios of branching fractions, and in comparison with the results of other theoretical approaches. In Sec. 5 we summarize our results. In the Appendix, we briefly describe the RIQ model framework, the wave packet representation of the participating meson states and momentum probability amplitude of the constituent quark and antiquark in the meson bound state.
		
	\section{General formalism and kinematics}
	The general formalism and kinematics for exclusive semileptonic $B_c$-decays: $B_c\to \eta_c/J/\psi l\nu_l$ and $B_c\to D/D^*l\nu_l$ are described elaborately in Ref.\cite{A41,A71}, from which we quote here a few important expressions related to leptonic and hadronic parts of the invariant decay amplitude:	
	\begin{equation}
		{\cal M}(p,k,k_l,k_\nu)={\frac{\cal G_F}{\sqrt{2}}}V_{bq^{'}}{\cal H}_\mu(p,k) {\cal L}^\mu(k_l,k_\nu), 
	\end{equation}

\noindent where ${\cal L}^\mu$ and ${\cal H}_\mu$ are leptonic and hadronic parts, respectively:
	\begin{eqnarray}
		&&{\cal L}^\mu(k_l,k_\nu)=\bar u(\vec{k_l})\gamma ^\mu (1-\gamma_5)v(\vec{k_\nu})\nonumber\\
		&&{\cal H}_\mu(p,k) =\langle X(\vec{k},S_X)|{J^h_\mu(0)}|B_c(\vec{p},S_{B_c})\rangle
	\end{eqnarray} 
	
	\noindent Here $J^h_\mu=V_\mu-A_\mu$ is the vector-axial vector current; $q^{'}=c,u$. We take $(p,k)$ as four momenta of the parent($B_c$) and daughter($X$) meson with their respective spin state: $S_{B_c}$ and $S_X$ and mass: M and m. $k_l$ and $k_\nu$ are the four momenta of lepton and lepton neutrino, respectively. $q=p-k=k_l+k_\nu$ represents the four-momentum transfer.	\\
	The hadronic amplitudes are covariantly expanded in terms of Lorentz invariant form factors. For $(0^- \to 0^-)$ type transitions, the expansion is
		\begin{equation}
		{\cal H}_\mu(B_c\to(\bar{c}c/\bar{u}c)_{S=0}) =(p+k)_\mu F_+(q^2)+q_\mu F_-(q^2),
	\end{equation}
	and for $(0^-\to 1^-)$ type transitions, its covariant expansion is
		\begin{eqnarray}
		{\cal H}_\mu(B_c\to(\bar{c}c/\bar{u}c)_{S=1})=&&{\frac{1}{(M+m)}}\epsilon ^{\sigma ^\dagger}\nonumber\\
		&&\Big\{g_{\mu\sigma}(p+k)q A_0(q^2)\nonumber\\
		&&+(p+k)_\mu(p+k)_\sigma A_+(q^2)\nonumber\\
		&&+q_\mu(p+k)_\sigma A_-(q^2)\nonumber\\&&+i\epsilon_{\mu\sigma\alpha\beta}(p+k)^\alpha q^\beta V(q^2)\Big\} 
	\end{eqnarray}
	
	\noindent The angular decay distribution differential in the momentum transfer squared $q^2$ is obtained in the form 
	\begin{equation}
		\frac{d\Gamma}{dq^2dcos\theta}	={\frac{{\cal G}_F}{(2\pi)^3}}|V_{bq^{'}}|^2\frac{(q^2-m_l^2)^2}{8M^2 q^2}|\vec{k}|{\cal L}^{\mu \sigma}{\cal H}_{\mu \sigma}
	\end{equation}
	Here ${\cal L}^{\mu \sigma}$ and ${\cal H}_{\mu \sigma}$ are the lepton and hadron tensor, respectively; $m_l$ is the mass of charged lepton.
	In our normalization, the lepton tensor ${\cal L}^{\mu\sigma}$ is found to be 
	\begin{eqnarray}
		{\cal L}^{\mu\sigma}_{\mp}=&&8\Bigg\{k^{\mu}_l k^{\sigma}_{\nu}+k^{\sigma}_l k^{\mu}_{\nu}-g^{\mu \sigma}\Bigg({\frac{q^2-m^2_l}{2}}\Bigg)\nonumber\\&&\pm i{\epsilon^{\mu \sigma\alpha \beta}}k_{l_\alpha} k_{\nu_\beta}\Bigg\}
	\end{eqnarray}

	\noindent It is convenient to express physical observables such as the hadron tensor on a helicity basis in which the helicity form factors are obtained in terms of the Lorentz invariant form factors. While doing the Lorentz contraction in Eq.(5) with the helicity amplitudes, one considers four covariant helicity  projections $\epsilon^{\mu}(m)$ with $m=+,-,0$ and $m=t$ for spin 1 and spin 0 part, respectively, of the $W_{off-shell}$ involved in the decay process. In our attempt to study the lepton mass effects in the semileptonic decay modes, we need to consider the time component of the polarization $\epsilon^{\mu}(m=t)$ in addition to its other three components with $m=+,-,0$.\\
	 Using orthogonality and the completeness relations satisfied by helicity projections, the Lorentz contraction in Eq.(5) can be written as
	
	\begin{eqnarray}
		{\cal L}^{\mu \sigma}{\cal H}_{\mu\sigma}               =&&{\cal L}_{\mu'\sigma'}g^{\mu'\mu}g^{\sigma'\sigma}{\cal H}_{\mu\sigma}\nonumber\\
		=&&{L(m,n)}{g_{mm'}}{g_{nn'}}H({m'}{n'})
	\end{eqnarray} 
	with $g_{mn}=dia(+,-,-,-)$.	Here the lepton and hadron tensors are introduced in the space of helicity components as:
	\begin{eqnarray}
		L(m,n)={\epsilon^{\mu}}(m){\epsilon^{\sigma ^\dagger}}(n){\cal L}_{\mu \nu}\nonumber\\
		H(m,n)={\epsilon^{\mu^\dagger}}(m){\epsilon^{\sigma}(n)}{\cal H}_{\mu \nu}
	\end{eqnarray}
	For the sake of convenience, we consider two frames of reference: i) the $\bar{l}\nu$ or $l\bar{\nu}$ center-of-mass frame and ii) the parent $B_c$-meson rest frame. We evaluate lepton tensor $L(m,n)$ in the  $\bar{l}\nu$ or $l\bar{\nu}$ c.m. frame and hadron tensor $H(m,n)$ in the $B_c$-rest frame.
	
	\noindent The space-and-time components of the four momenta: $p^\mu,k^\mu,q^\mu$, and that of the polarization vectors: $\epsilon^\mu(\pm)$, $\epsilon^{\mu}(0)$, and $\epsilon^\mu(t)$ are also specified in the $B_c$-rest frame from which the helicity components of the hadronic tensors are obtained in terms of Lorentz invariant form factors.
	For $B_c\to (\bar{c}c/\bar{u}c)_{s=0}$ transition, they are obtained as
	\begin{eqnarray}
		{H_t}=&&\frac{1}{\sqrt{q^2}}\biggl\{(p+k).(p-k){F_+}+{q^2}{F_-}\biggr\}\nonumber\\
		{H_\pm}=&&0\\
		{H_0}=&&\frac{2M\vert\vec{k}\vert}{\sqrt{q^2}}{F_+}\nonumber
	\end{eqnarray}
	
	\noindent and for $B_c\to (\bar{c}c/\bar{u}c)_{s=1}$ transitions, the relations between helicity form factors and invariant form factors are obtained in the form
	\begin{eqnarray}
		{H_t}=&&\frac{1}{(M+m)}\frac{M{\vert \vec{k}\vert}}{m\sqrt{q^2}}\{(p+k) {.} q\ ({-A_0}+{A_+})+{{q^2}{A_-}}\}\nonumber\\
		{H_\pm}=&&\frac{1}{(M+m)}\big\{-(p+k) {.}q\ {A_0}\mp 2M\vert\vec{k}\vert V\big\}\\
		{H_0}=&&\frac{1}{(M+m)}\frac{1}{2m\sqrt{q^2}}\Big\{-(p+k){.} q ({M^2}-{m^2}-{q^2}){A_0}\nonumber\\
		&&+4{M^2}\vert\vec{k}\vert^2 {A_+}\Big\} \nonumber
	\end{eqnarray}  
Using the space- and time components of the four momenta: $(q^{\mu},k_l^{\mu}, k_{\nu}^{\mu})$ and helicity projections: $\epsilon^{\mu}(m=+,-,0;t)$ in the $(l\bar{\nu})$-c.m. frame, it is straight forward to calculate the helicity representation $L(m,n)$ of the lepton tensor (8).	
	In the present analysis, we do not consider the azimuthal $\chi$ distribution of the lepton pair and therefore integrate over the azimuthal angle dependence of the lepton tensor. The differential $(q^2,\cos\theta)$ distribution is then obtained in the form:
	\begin{eqnarray}
		\frac{d\Gamma}{d{q^2}\cos\theta}=&&\frac{3}{8}(1+\cos^2\theta)\frac{d\Gamma_U}{dq^2}+\frac{3}{4}\sin^2\theta.\frac{d\Gamma_L}{dq^2}\mp \frac{3}{4} \cos\theta \frac{d\Gamma_P}{dq^2}\nonumber\\
		&&+\frac{3}{4}\sin^2\theta\frac{d{\tilde{\Gamma}_U}}{dq^2}+\frac{3}{2}\cos^2\theta\frac{d{\tilde{\Gamma}_L}}{dq^2}\nonumber\\
		&&+\frac{1}{2}\frac{d{\tilde{\Gamma}}_S}{dq^2}+3\cos\theta\frac{d{\tilde{\Gamma}}_{SL}}{dq^2}
	\end{eqnarray} 
	The upper and lower signs associated with the parity-violating term in Eq.(11) refer to two cases: $l^-\bar{\nu}$ and $l^+{\nu}$, respectively. Out of seven terms in the r.h.s of Eq.(11), four terms identified as tilde rates $\tilde{\Gamma}_i$ are linked with the lepton mass, and other terms identified as $\Gamma_i$ are lepton mass-independent. Both are related via a flip factor $\frac{m_{l}^2}{2q^2}$ as:
	\begin{equation}
		\frac{d{\tilde{\Gamma}}_i}{dq^2}=\frac{m_l^2}{2q^2}\frac{d{\Gamma_i}}{dq^2}
	\end{equation}
	The tilde rates do not contribute to the decay rate in the vanishing lepton mass limit. They can be neglected for $e$ and $\mu$ modes but they are expected to provide a sizeable contribution to the $\tau$-modes. Therefore the tilde rates are crucial in evaluating the lepton mass effects in the semileptonic decay modes. The differential partial helicity rates $\frac{d\Gamma_i}{dq^2}$ are defined by
	\begin{equation}
		\frac{d\Gamma_i}{dq^2}=\frac{{\cal G}_f^2}{(2\pi)^3}{\vert V_{bq^{'}}\vert}^2 \frac{({q^2}-{m_e^2})^2}{12{M^2}q^2}|\vec{k}|H_i
	\end{equation}
	Here $H_i(i=U,L,P,S,SL)$ represents a standard set of helicity structure function given by linear combinations of helicity components of hadron tensor $H(m,n)=H_m H_n^\dagger$:
	
	\begin{eqnarray}
		{H_U}=&&Re({H_+}{H_+^\dagger})+Re({H_-}{H_-^\dagger}) \ \  :Unpolarized-transversed\nonumber\\
		{H_L}=&& Re({H_0}{H_0^\dagger})\ \ \ \ \ \ \ \ \ \ \ \ \ \ \ \ \ \ \ \ \ \ \ \ \ \  :Longitudinal\nonumber\\
		{H_P}=&&Re({H_+}{H_+^\dagger})-Re({H_-}{H_-^\dagger})\ \ \ :Parity-odd\nonumber\\
		{H_S}=&&3Re({H_t}{H_t^\dagger})\ \ \ \ \ \ \ \ \ \ \ \ \ \ \ \ \ \ \ \ \ \ \ \ \ :Scalar\nonumber\\
		{H_{SL}}=&&Re({H_t}{H_0^\dagger})\ 
		\ \ \ \ \ \ \ \ \ \ \ \ :Scalar-Longitudinal\ Interference\nonumber
	\end{eqnarray}
	Here we assume that the helicity amplitudes are real since the available $q^2$- range: ($q^2\le(M-m)^2$) is below the physical threshold $q^2= (M+m)^2$. Therefore we drop the angular terms that are multiplied by coefficients $Im(H_iH_j^*),i\ne j^*$. Then integrating over $\cos\theta$ one gets the differential $q^2$ distribution and finally integrating over $q^2$, one obtains the total decay rate $\Gamma$ as the sum of the partial decay rates :
	$\Gamma_i=(i=U,L,P)$ and $\tilde{\Gamma}_i(i=U,L,S,SL)$.\\
	
	A quantity of interest is the forward-backward asymmetry $A_{FB}$ of the lepton in the $(l\bar{\nu})$ c.m. frame which is defined as\\
	\begin{equation}
		A_{FB}=\frac{3}{4}\biggl\{\frac{\pm P+4\tilde{SL}}{U+\tilde{U}+L+\tilde{L}+\tilde{S}}\biggr\}
	\end{equation}
	Another quantity of interest is the asymmetry parameter $\alpha^*$ which is defined by rewriting Eq.(11) in terms of its $\cos^2\theta^*$ dependence i.e $d\Gamma \propto 1+\alpha^* \cos ^2\theta^*$. The asymmetry parameter $\alpha^*$ which determines the transverse and longitudinal composition of the final state vector meson is given by: 
	\begin{eqnarray}
		\alpha^*=\frac{U+\tilde{U}-2(L+\tilde{L}+\tilde{S})}{U+\tilde{U}+2(L+\tilde{L}+\tilde{S})} 
	\end{eqnarray}
	We list our predictions on helicity rates $\Gamma_i$, $\tilde{\Gamma}_i$, $A_{FB}$, and $\alpha^*$ in Sec. IV.
		\section{Transition amplitude and weak decay form factors}
	\subsection{Transition amplitude}
	The decay process physically occurs in the momentum eigenstate of the participating mesons. Therefore its field-theoretic description requires meson bound-states to be represented by appropriate momentum wave packets with suitable momentum and spin distribution between the constituent quark-antiquark pair in the corresponding meson core. In the relativistic independent particle picture of the RIQ model, the constituent quark antiquark are assumed to be in the state of independent and relativistic motion inside the meson bound-state: $|B_c(\vec{p}_b,\vec{p}_c)\rangle $, for example, with the momentum $\vec{p}_b$ and $\vec{p}_c$, respectively. In this model, the momentum probability amplitude of the constituent quark and antiquark in the participating meson ground states are obtained by taking the momentum projections of corresponding quark orbitals. The corresponding expressions (A.6) are shown in the Appendix. In the present context, one needs to extract equivalent model expressions for the momentum probability amplitude of constituent quark and antiquark inside the charmonium and charm mesons in their radially excited $2S$ and $3S$ states. For this, the quark orbitals for radially excited meson states are obtained by using the relevant Dirac quantum numbers and associated Laguerre polynomial in the general expression(A1) in the Appendix. Thereafter we take the momentum projection of the relevant quark orbitals and obtain the momentum probability amplitude of the quark and antiquark in the $S$-wave ($2S$ and $3S$) charmonium and charm meson states, respectively, in the form;
	\begin{eqnarray}
		G_b(\vec p_b)&&={{i\pi {\cal N}_b}\over {2\alpha _b}}
		\sqrt {{(E_{p_b}+m_b)}\over {E_{p_b}}} {(E_{p_b}+E_b)\over{(E_b+m_b)}}\nonumber\\
		&&({\vec {p_b}^2\over {2\alpha _b}}-{3\over 2})
		\exp {(-{{\vec p_b}^2\over {4\alpha_b}})}\nonumber\\
		{\tilde G}_c(\vec p_c)&&={{i\pi {\cal N}_c}\over {2\alpha _c}}
		\sqrt {{(E_{p_c}+m_c)}\over {E_{p_c}}}
		{(E_{p_c}+E_c)\over {(E_c+m_c)}}\nonumber\\
		&&({\vec {p_c}^2\over {2\alpha _c}}-{3\over 2})
		\exp {(-{{\vec p_c}^2\over {4\alpha_c}})}
	\end{eqnarray}  
	and                 		
	\begin{eqnarray}  
		G_b(\vec p_b)&&={{i\pi {\cal N}_b}\over {2\alpha _b}}
		\sqrt {{(E_{p_b}+m_b)}\over {E_{p_b}}} {(E_{p_b}+E_b)\over{(E_b+m_b)}}\nonumber\\
		&&({\vec {p_b}^4\over {8 {\alpha _b}^2}}-{{5{\vec p_b}^2}\over {4\alpha_b}}+{{15\over 8}})
		\exp {(-{{\vec p_b}^2\over {4\alpha_b}})}\nonumber\\
		{\tilde G}_c(\vec p_c)&&={{i\pi {\cal N}_c}\over {2\alpha _c}}
		\sqrt {{(E_{p_c}+m_c)}\over {E_{p_c}}}
		{(E_{p_c}+E_c)\over {(E_c+m_c)}}\nonumber\\
		&&({\vec {p_c}^4\over {8 {\alpha _c}^2}}-{{5{\vec p_c}^2}\over {4\alpha_c}}+{{15\over 8}})
		\exp {(-{{\vec p_c}^2\over {4\alpha_c}})}
	\end{eqnarray} 
	The binding energy of constituent quark and antiquark in the radially excited meson states are also obtained by solving the appropriate cubic equations for the binding energy condition (A5) shown in the Appendix. After specifying the momentum probability amplitudes of constituent quarks and their binding energies, the radially excited final meson states, involved in the semileptonic $B_c$-meson decays, are duly constructed which can be used to calculate the $S$-matrix elements for the decay process.\\
	The internal dynamics at the constituent level, responsible for physically observable meson-level decay processes, are described by an otherwise unbound quark and antiquark using the usual Feynman technique. The constituent level $S$-matrix element $S_{fi}^{b\to c,u}$, obtained from the Feynman diagram, when operated upon by the bag like operator $\hat{\Lambda}$ used in the RIQ model  yields meson-level $S$-matrix in the form
	
	\begin{equation}
		\hat{\Lambda}S_{fi}^{b\to c/u}\longrightarrow S_{fi}^{{B_c}\to(\bar{c}c/\bar{u}c)system}
	\end{equation}
	\begin{figure}[!hbt]
	\includegraphics{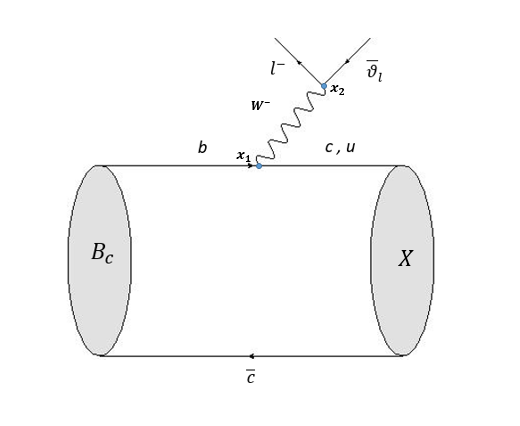}\caption{Leading order diagram of semileptonic decay of $B_c$ meson.}
		\label{FD}
	\end{figure}
	
	Using the wave packet representation of the participating meson states, the $S$-matrix element for the decay process is obtained in the standard form 
	\begin{eqnarray}
		S_{fi}=&&(2\pi)^4\delta^{(4)}(p-k-k_l-k_\nu)\nonumber\\
		&&(-{\cal M}_{fi})\frac{1}{\sqrt{(2\pi)^32E_{B_c}}}\Pi_f \frac{1}{\sqrt{2E_f(2\pi)^3}} 
	\end{eqnarray}
	 In the $B_c$-rest frame the hadronic amplitude ${\cal H}_{\mu}$ is
	\begin{eqnarray}
		{\cal H}_\mu=&&\sqrt{\frac{4ME_k}{N_{B_c}(0)N_X(\vec{k})}}\int\frac{d^3p_b}{\sqrt{2E_{p_b}2E_{k+p_b}}}\nonumber\\
		&&{\cal G}_{B_c}(\vec{p_b},-\vec{p_b}){\cal G}_X(\vec{k}+\vec{p_b},-\vec{p_b})\langle S_X\vert J_\mu^h(0)\vert S_{B_c}\rangle  
	\end{eqnarray}
	where $E_{p_b}$ and $E_{p_b+k}$ stand for the energy of the non-spectator quark of the parent and daughter meson, respectively, and $\langle S_X\vert J_\mu^h(0)\vert S_{B_c}\rangle $ represents symbolically the spin matrix elements of vector-axial vector current.
	
	\subsection{Weak decay form factors}
	For $0^-\to 0^-$ transitions, the axial vector current does not contribute. The spin matrix elements corresponding to the non-vanishing vector current parts are obtained in the form:
	\begin{eqnarray}
		\langle S_X(\vec{k})\vert V_0\vert S_{B_c}(0)\rangle =\frac{(E_{p_b}+m_b)(E_{p_{c/u}}+m_{c/u})+|\vec{p_b}|^2}{\sqrt{(E_{p_b}+m_b)(E_{p_{c/u}}+m_{c/u})}}\\
		\langle S_X(\vec{k})\vert V_i\vert S_{B_c}(0)\rangle =\frac{(E_{p_b}+m_b)k_i}{\sqrt{(E_{p_b}+m_b)(E_{p_b+k}+m_{c/u})}}
	\end{eqnarray}	  
	With the above spin matrix elements, the expressions for hadronic amplitudes (20) are compared with corresponding expressions  in Eq.(3) yielding the form factors $f_+$ and $f_-$ as
	\begin{eqnarray}
		f_\pm (q^2)=&&\frac{1}{2M}\sqrt{\frac{ME_k}{N_{B_c}(0)N_X(\vec{k})}}\nonumber\\ && \int d\vec{p_b}{\cal G}_{B_c}(\vec{p_b},-\vec{p_b}){\cal G}_X(\vec{k}+\vec{p_b},-\vec{p_b})\nonumber\\ && \frac{(E_{o_b}+m_b)(E_{p_{c/u}}+m_{c/u})+|\vec{p_b}|^2\pm(E_{p_b}+m_b)(M\mp E_k)}{E_{p_b}E_{p_{c/u}}(E_{p_b}+m_b)(E_{p_{c/u}}+m_{c/u})}	
	\end{eqnarray}
	For $(0^-\to 1^-)$ transitions, the spin matrix elements corresponding to the vector and axial-vector current are found separately as:
	\begin{eqnarray}
		\langle S_X(\vec{k},\hat{\epsilon^*})\vert V_0\vert S_{B_c}(0)\rangle =&&0\\
		\langle S_X(\vec{k},\hat{\epsilon^*})\vert V_i\vert S_{B_c}(0)\rangle=&&\frac{i(E_{p_b}+m_b)(\hat{\epsilon}^*\times \vec{k})_i}{\sqrt{(E_{p_b}+m_b)(E_{p_b+k}+m_{c/u})}}\\
		\langle S_X(\vec{k},\hat{\epsilon^*})\vert A_i\vert S_{B_c}(0)\rangle=&&\frac{(E_{p_b}+m_b)(E_{p_b+k}+m_{c/u})-\frac{|\vec{p_b}|^2}{3}}{\sqrt{(E_{p_b}+m_b)(E_{p_b+k}+m_{c/u})}}\\
		\langle S_X(\vec{k},\hat{\epsilon^*})\vert A_0\vert S_{B_c}(0)\rangle=&&\frac{-(E_{p_b}+m_b)(\hat{\epsilon}^*. \vec{k})}{\sqrt{(E_{p_b}+m_b)(E_{p_b+k}+m_{c/u})}}
	\end{eqnarray}
	With the spin matrix elements (24-27), the expressions for hadronic amplitudes (20) are compared with corresponding expressions in Eq.(4). Then the model expressions of the form factors: $V(q^2),A_0(q^2),A_+(q^2)$ and $A_-(q^2)$ are obtained in the form:
	\begin{eqnarray}
		V(q^2)=&&\frac{M+m}{2M}\sqrt{\frac{ME_k}{N_{B_c}(0)N_X(\vec{k})}}\int d\vec{p_b}{\cal G}_{B_c}(\vec{p_b},-\vec{p_b})\nonumber\\ &&{\cal G}_X(\vec{k}+\vec{p_b},-\vec{p_b})\times \sqrt{\frac{(E_{p_b}+m_b)}{E_{p_b}E_{p_{c/u}}(E_{p_{c/u}}+m_{c/u})}}	\\
		A_0(q^2)=&&\frac{1}{(M-m)}\sqrt{\frac{Mm}{N_{B_c}(0)N_X(\vec{k})}}\nonumber\\ &&\times\int d\vec{p_b}{\cal G}_{B_c}(\vec{p_b},-\vec{p_b}){\cal G}_X(\vec{k}+\vec{p_b},-\vec{p_b})\nonumber\\ &&\times \frac{(E_{p_b}+m_b)(E^0_{p_{c/u}}+m_{c/u})-\frac{|\vec{p_b}|^2}{3}}{\sqrt{E_{p_b}E_{p_{c/u}}(E_{p_b}+m_b)(E_{p_{c/u}}+m_{c/u})}}
	\end{eqnarray}
	with $E^0_{p_{c/u}}=\sqrt{|\vec{p}_{c/u}|^2+m^2_{c/u}}$
	and\begin{equation}
		A_\pm(q^2)=\frac{-E_k(M+m)}{2M(M+2E_k)}\Bigg[T\mp \frac{3(M\mp E_k)}{(E_k^2-m^2)}\big\{I-A_0(M-m)\big\}\Bigg]
	\end{equation}
	\noindent with $T=J-(\frac{M-m}{E_k})A_0$, where
	\begin{eqnarray}
		J=&&\sqrt{\frac{ME_k}{N_{B_c}(0)N_X(\vec{k})}}\int d\vec{p_b}{\cal G}_{B_c}(\vec{p_b},-\vec{p_b}){\cal G}_X(\vec{k}+\vec{p_b},-\vec{p_b})\nonumber\\
		&&\times \sqrt{\frac{(E_{p_b}+m_b)}{E_{p_b}E_{p_{c/u}}(E_{p_{c/u}}+m_{c/u})}}\\
		I=&&\sqrt{\frac{ME_k}{N_{B_c}(0)N_X(\vec{k})}}\int d\vec{p_b}{\cal G}_{B_c}(\vec{p_b},-\vec{p_b}){\cal G}_X(\vec{k}+\vec{p_b},-\vec{p_b})\nonumber\\
		&&\times \biggl\{\frac{(E_{p_b}+m_b)(E^0_{p_{c/u}}+m_{c/u})-\frac{|\vec{p}_b|^2}{3}}{\sqrt{E_{p_b}E^0_{p_{c/u}}(E_{p_b}+m_b)(E^0_{p_{c/u}}+m_{c/u})}}\biggr\}
	\end{eqnarray}
	With the form factors thus obtained in terms of model quantities, the helicity amplitudes and hence the decay rates for semileptonic $B_c$-meson decays to radially excited $(2S$ and $3S) $ charmonium and charm meson states are evaluated and our predictions are listed in Sec.4.
	
	\section{Numerical results and discussion}
	In this section, we present our numerical results on the semileptonic $B_c$ decay modes to $S$-wave charmonium and charm meson states. For numerical calculation, we need the model parameters $(a,\ V_0)$, quark masses $m_q$, and quark binding energies $E_q$, which have already been fixed from hadron spectroscopy by fitting the data of heavy and heavy-light flavored mesons in their ground state as 
	\begin{eqnarray}
		(a,\ V_0)=&&(0.017166\ GeV^3, -0.1375\ GeV)\nonumber\\
		(m_b,\ m_c,\ m_u)=&&(4.77659, 1.49276,0.07875)GeV\nonumber\\
		(E_b,\ E_c,\ E_u)=&&(4.76633, 1.57951, 0.47125)GeV
	\end{eqnarray}  
	With this set of input parameters, wide-ranging hadronic phenomena have been described earlier in the framework of the RIQ model. However, for a description of the decay process, in the present context, involving radially excited final meson states, the constituent quarks in the meson bound states are expected to have higher binding energies compared to the binding energies in their ground states. We solve the cubic equation representing the binding energy condition (A5) for respective constituent quarks and obtain the binding energies of the quarks $(c, u)$ in the radially excited $2S$ and $3S$ states of the $(\bar{c}c)$ and $(\bar{u}c)$ systems as:
	\begin{equation}
		(E_c, E_u)_{2S}=(1.97015, 0.96221) GeV
	\end{equation}   
	\begin{equation}
		(E_c,E_u)_{3S}=(2.22478,1.29356) GeV
	\end{equation}
	For the CKM-parameters and $B_c$-lifetime, we take their central values from the Particle Data Group (2020) \cite{A73} as:
	\begin{eqnarray}
		(V_{bc},V_{bu})=&&(0.041,0.00382)\nonumber\\
		\tau_{B_c}=&&0.510\ ps
	\end{eqnarray}
	For the masses of participating mesons, taken as phenomenological inputs in the present calculation, we take their central values of the observed data from \cite{A73,A74,A75,A76}. In the absence of observed data in the charmonium and charm meson sector, we take the corresponding predicted values from established theoretical approaches \cite{A77,A78,A79}. Accordingly, the updated meson masses used in our numerical analysis are listed in Table \ref{tab1}. 
	
	 \begin{table}[!hbt]
		\renewcommand{\arraystretch}{1}
		\centering
		\setlength\tabcolsep{5pt}
		\caption{The masses of the participating mesons}
		\label{tab1}
		\begin{tabular}{lll}
			\hline
			\hline Particle &Mass(MeV) &Reference  \\
			
			\hline$B_c$&6274.47&\cite{A73}\\
			$\eta_c(2S)$&3637&\cite{A73}\\
			$\psi(2S)$&3686.1& \cite{A73}\\
			$D^0(2S)$&2518& \cite{A74}\\
			$D^{0*}(2S)$&2681.1& \cite{A75}\\
			$\psi(3S)$&4039.1 & \cite{A76} \\
			$\eta_c(3S)$&4007 &\cite{A77} \\
			$D^0(3S)$&3068  &\cite{A78} \\
			$D^{0*}(3S)$&3110 &\cite{A79} \\
			\hline
		\end{tabular}
	\end{table}
	With these input parameters, the Lorentz invariant form factors: $(F_+,\  F_-;\  A_0,\ A_+,\ A_-,\ V)$ representing decay amplitudes can be calculated from the overlapping integral of participating meson wave functions. We first study the $q^2$-dependence of the invariant form factors in the allowed kinematic range. Since our main purpose is to study the lepton mass effect in semileptonic decay of $B_c$ meson, we  plot $q^2$-dependence of the form factors in Figs. 2 and 3 for decays in their $e^-,\mu^-$ and $\tau^-$-modes. We find that the behavior of form factors in $e^-$ and $\mu ^-$-mode overlap in the entire kinematic range: $0\le q^2\le q^2_{max}$. This is because of an insignificant change in the phase space  boundary going from $e^-$ to $\mu^-$ mode. As one can see here the maximal lepton energy shift $\frac{m^2_\mu-m^2_e}{2M}$ is invisible at the usual scale of the plot. On the other hand for the decays in the $\tau^-$ mode, the relevant form factors behave differently throughout the accessible kinematic range of $q^2_{min}\le q^2\le q^2_{max}$, where $q^2_{min}$ is +ve and away from $q^2\to 0$. The $\tau$-phase space, as compared to the $e^-$ and $\mu ^-$ cases is considerably reduced and shifted to a large $q^2$-region. In the present analysis, we shall therefore consider decays in their $e^-$ and $\tau ^-$ modes only for evaluating the lepton mass effects on the semileptonic $B_c$ meson decays.
	
	\begin{figure}[!hbt]
		\includegraphics[width=0.4\textwidth]{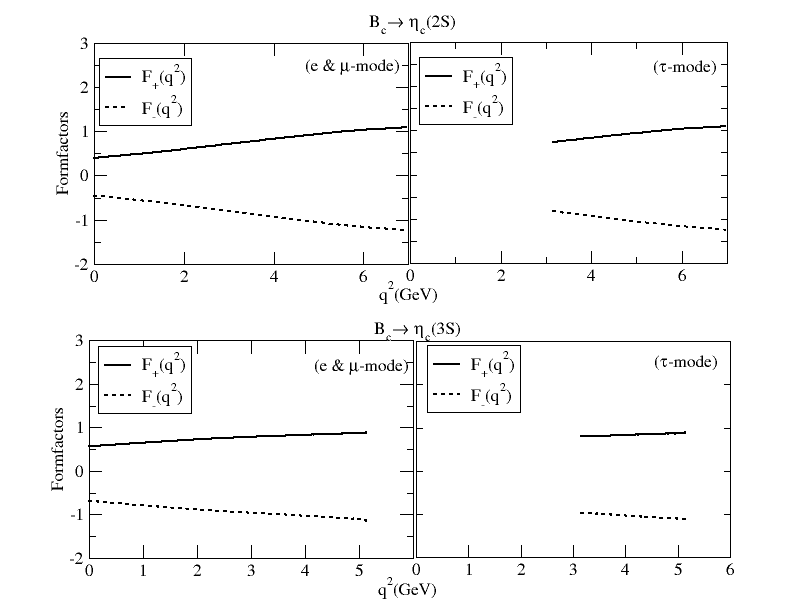}
		\includegraphics[width=0.4\textwidth]{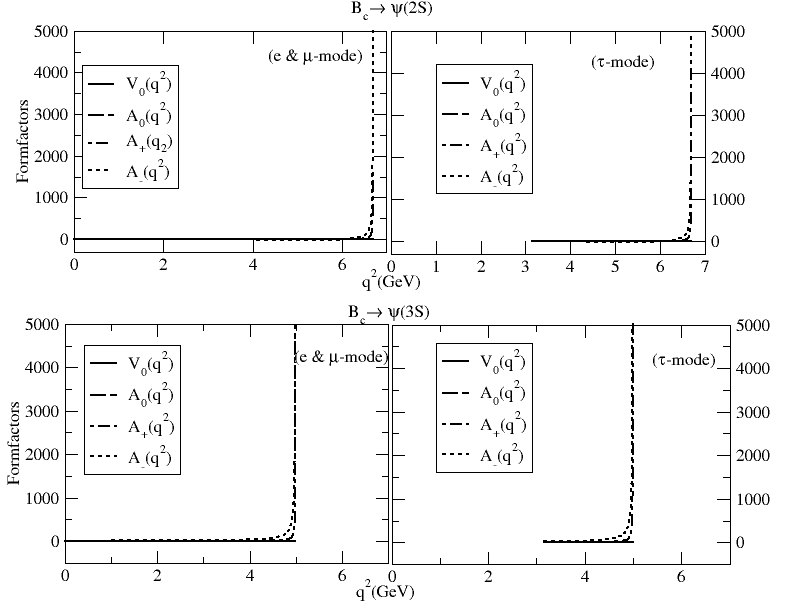}
		\caption{The $q^2$-dependence of invariant form factors for  $B_c \to \eta_c(2S)/\psi(2S)$ and $B_c \to \eta_c(3S)/\psi(3S)$ decays.}
	\end{figure}
	
	\begin{figure}[!hbt]
		\includegraphics[width=0.4\textwidth]{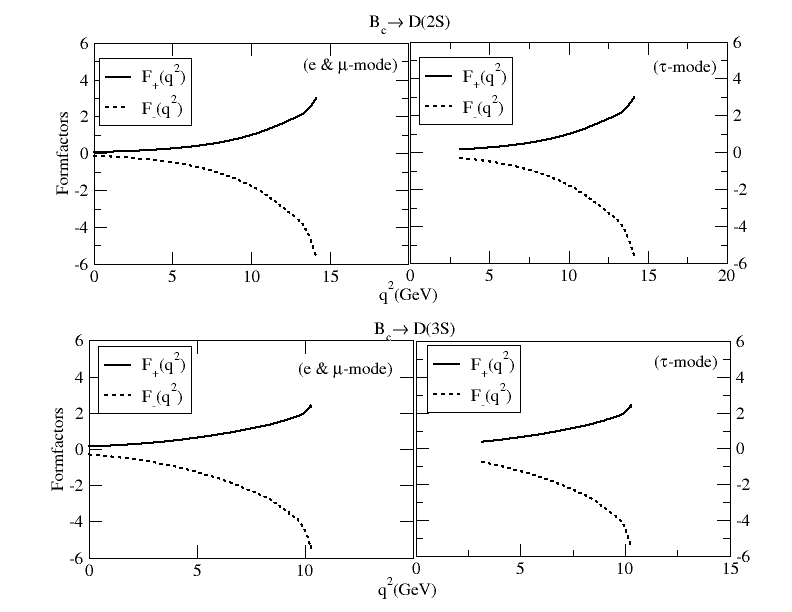}
		\includegraphics[width=0.4\textwidth]{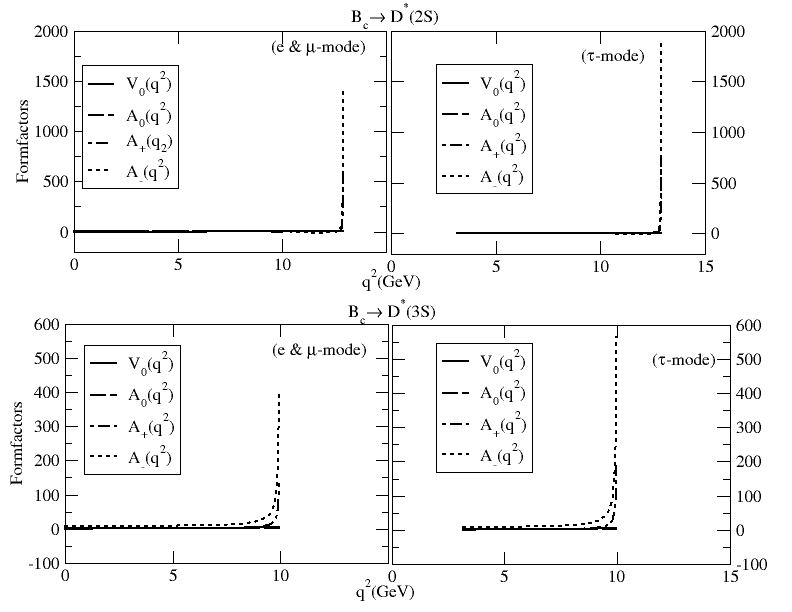}
		\caption{The $q^2$-dependence of invariant form factors for  $B_c \to D(2S)/D^*(2S)$ and $B_c \to D(3S)/D^*(3S)$ decays.}
	\end{figure}
	As mentioned earlier it is convenient to calculate decay amplitudes in the helicity basis in which the partial helicity rate and total decay rates are expressed in terms of helicity form factors. Using the helicity form factors it is straightforward to obtain partial helicity rates:$\frac{d\Gamma_i}{dq^2}$ (without flip) and $\frac{d\tilde{\Gamma}_i}{dq^2}$(with flip) for decay processes considered here in their $e^-$ and $\tau^-$ modes. For this we study the $q^2$-dependence  of the helicity form factors, partial helicity rates, and $q^2$-spectra of semileptonic decay rates, separately in their $e^-$ and $\tau ^-$ modes before evaluating helicity rates. We first rescale the helicity form factors according to
	\begin{equation}
		h_j=A(q^2)H_j,\ \ \ \ j=0,+,-\ \ \ (for\  no\ flip\ case)
	\end{equation}   
	and for flip cases
	\begin{eqnarray}
		\tilde{h}_j=&&\sqrt{\frac{m_l^2}{2q^2}}A(q^2)H_j\ ,\ \ \ \ \ j=0,+,-\nonumber\\
		\tilde{h}_t=&&\sqrt{\frac{m_l^2}{2q^2}}\sqrt{3}A(q^2)H_t
	\end{eqnarray}
	where \begin{equation}
		A(q^2)=\frac{{\cal G}_F}{4M}\big(\frac{q^2-m_l^2}{q^2}\big)\sqrt{\frac{|\vec{k}|q}{6\pi^3}}\ |V_{b,c/u}|
	\end{equation}
	and $\sqrt{\frac{m_l^2}{2q^2}}$: denotes a flip factor.	In terms of the rescaled  helicity form factors, the angle integrated differential $q^2$-rate is expressed as:
	\begin{equation}
		\frac{d\Gamma}{dq^2}=\sum_{0,+,-}|h_j|^2+\sum_{t,0,+,-}|\tilde{h}_j|^2 
	\end{equation}
	\begin{figure}[!hbt]
		\includegraphics[width=0.4\textwidth]{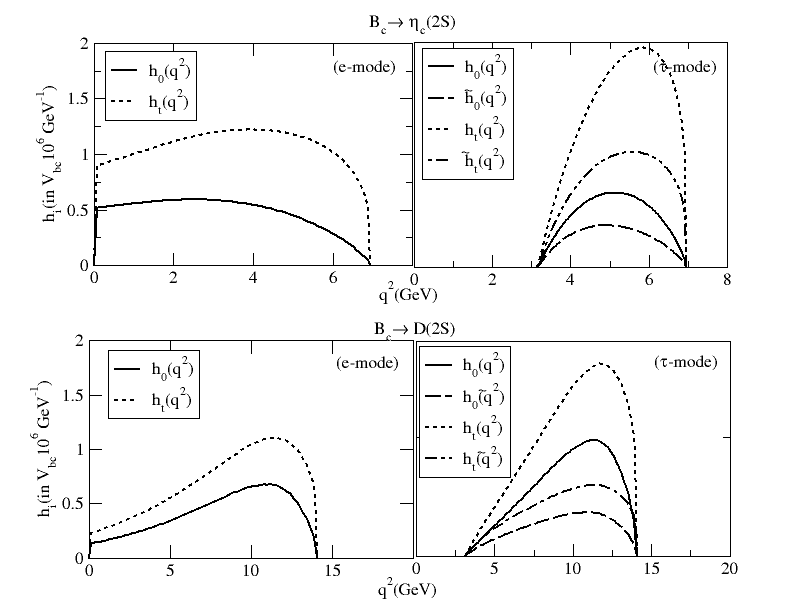}
		\includegraphics[width=0.4\textwidth]{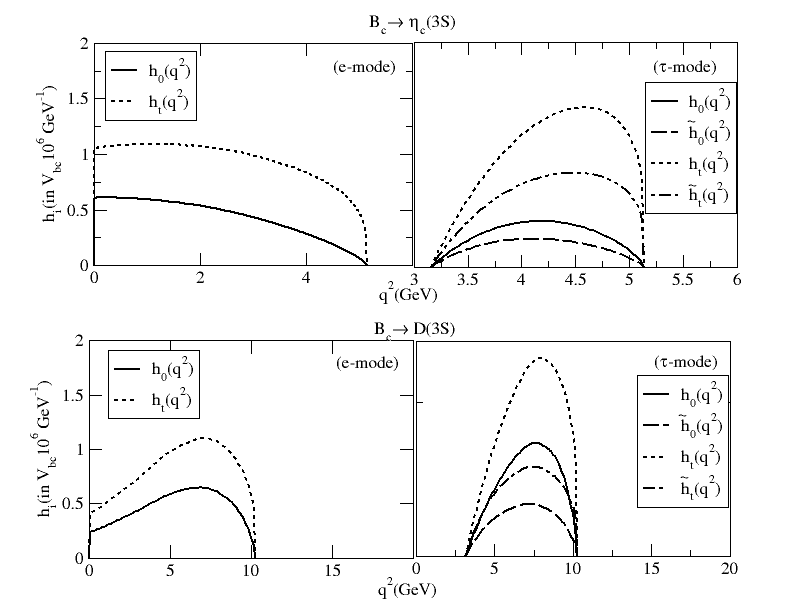}
		\caption{Reduced helicity amplitudes $h_j$ and $\tilde{h_j}$, $(j=t,+,-,0)$ as functions of $q^2$ for  $B_c \to \eta_c(2S)/D(2S)$ and $B_c \to \eta_c(3S)/D(3S)$ decays.}
	\end{figure}
	\begin{figure}[!hbt]
		\includegraphics[width=0.3\textwidth]{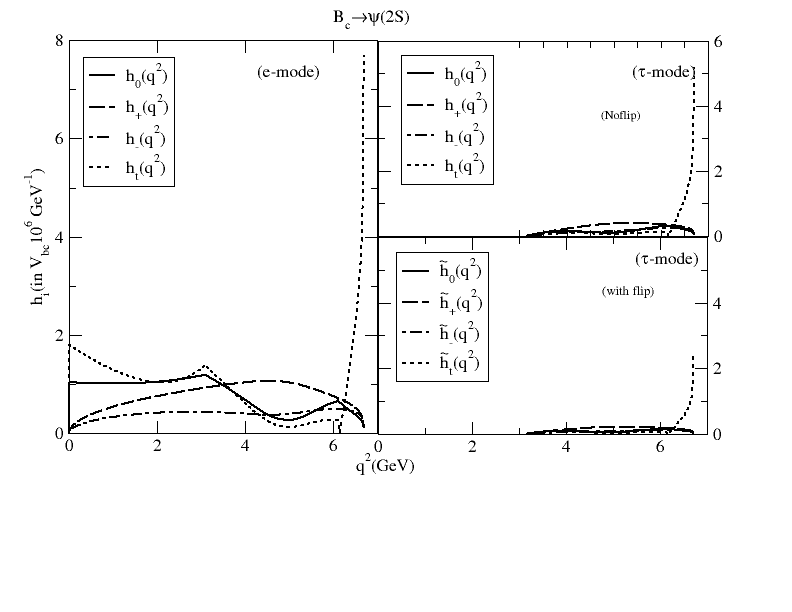}
		\includegraphics[width=0.3\textwidth]{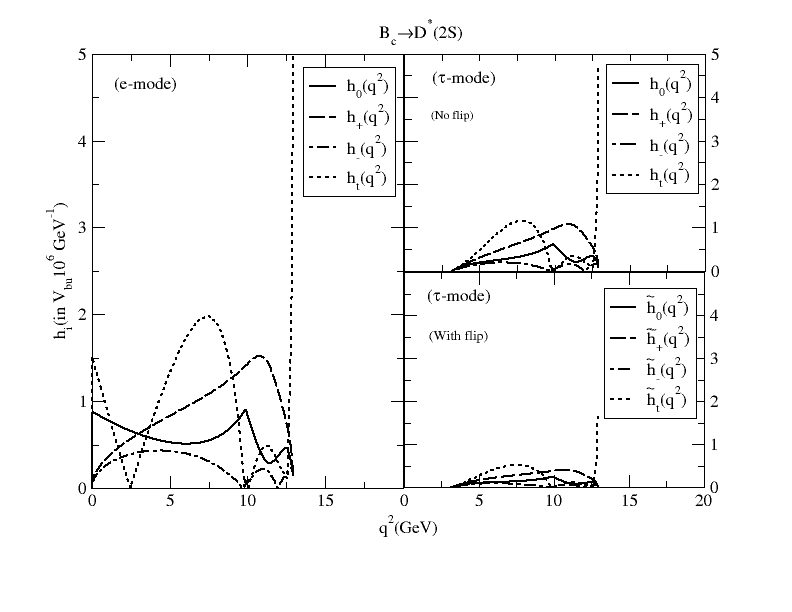}
		\includegraphics[width=0.3\textwidth]{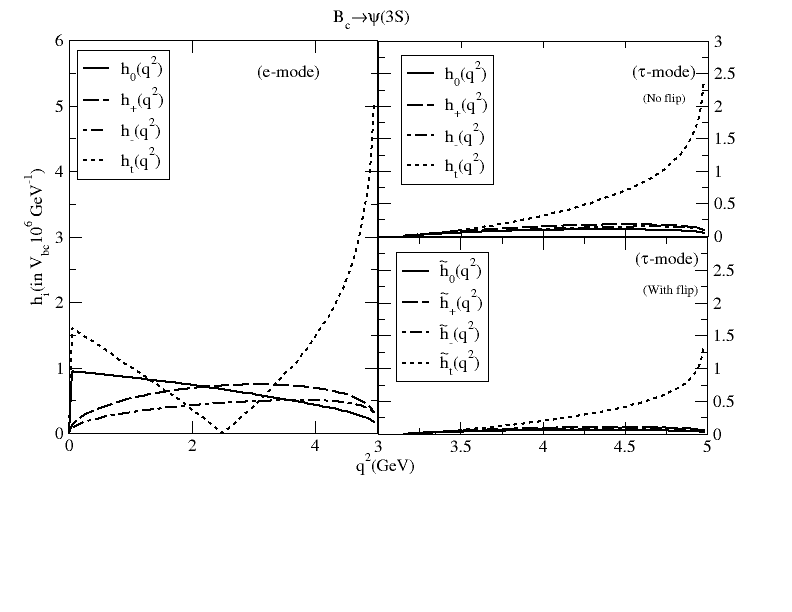}
		\includegraphics[width=0.3\textwidth]{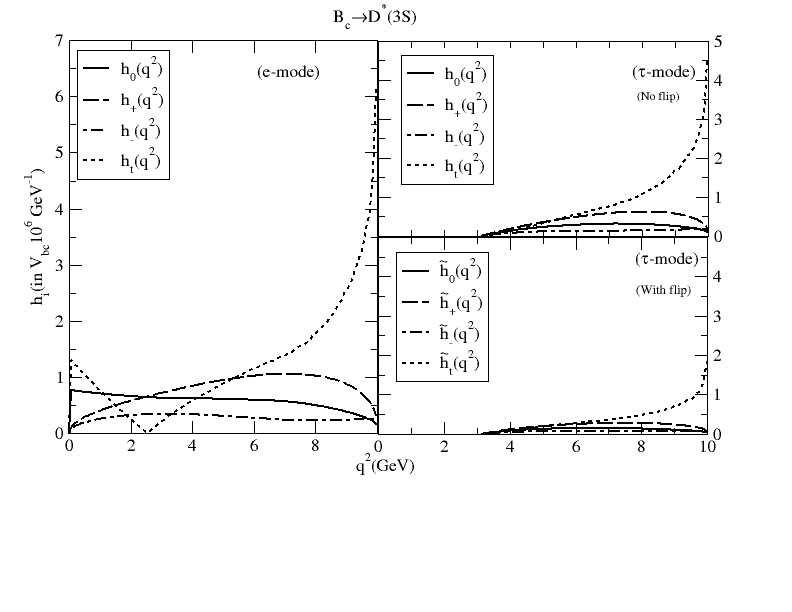}
		\caption{Reduced helicity amplitudes $h_j$ and $\tilde{h_j}$, $(j=t,+,-,0)$ as functions of $q^2$ for semileptonic decay rates for $B_c \to \psi(2S)/D^*(2S)$ and $B_c \to \psi(3S)/D^*(3S)$ decays.}
	\end{figure}

	In Fig.4 we plot the $q^2-$dependence of the rescaled helicity form factors $h_j$ and $\tilde{h}_j$ for $B_c\to \eta_c (2S,3S)$ and $B_c\to D(2S,3S)$ decays, respectively in their $e^-$ and $\tau^-$modes. We find that the longitudinal no-flip amplitude $h_0$ is reduced in the low $q^2$ region going from $e^-$ to $\tau^-$ mode. This reduction is due to the threshold like factor; $\frac{q^2-m^2_l}{q^2}$ appearing in the rescaled helicity amplitude. The longitudinal flip amplitude $\tilde{h}_0$ is further reduced by the flip factor $\sqrt{\frac{m_l^2}{2q^2}}$. The large value of the scalar flip amplitude $\tilde{h}_t$ is attributed to time-like (scalar-)current contribution which proceeds via an orbital $S-$wave, where there is no pseudo-threshold  factor to tamper the enhancement at large $q^2$ resulting from the time-like form factor in the helicity amplitude.\\
	
	In Fig.5 we plot the $q^2$-dependence of the rescaled helicity amplitudes for $B_c\to \psi(2S, 3S)$ and $B_c\to D^*(2S,3S)$ transitions in their $e^-$ and $\tau^-$ mode. We find a significant reduction of the longitudinal no-flip amplitude $h_0$ in these decay modes. Contrary to $B_c\to \eta_c(2S,3S)/D(2S,3S)$ decays we find all the flip amplitudes are generally small compared to corresponding no-flip amplitudes. This is because of the partial wave structure of the scalar-current contribution. The effects of timelike (scalar)-current on semileptonic decay in the $\tau^-$ mode are depicted in Figs.4 and 5. It may be noted here that the contributions of invariant form factors: $F_-(q^2)$ and $A_-(q^2)$ are dominant to determine the shape of the plot of flip helicity component $\tilde{h}_t$ over the allowed kinematic range of $q^2$. In $B_c\to \eta_c(2S,3S)/D(2S,3S)$ decay modes the contribution of $F_-(q^2)$ to the timelike helicity part is destructive for which the corresponding rescaled helicity amplitude $\tilde{h}_t$ increases when $F_-(q^2)$ is switched off as shown in Fig.4. Since $|\tilde{h}_t|^2$ contributes dominantly to the decay rate, an accurate determination of $B_c\to \eta_c(2S,3S)$ and $B_c\to D(2S,3S)$ decays in their $\tau ^-$-mode would help extract information on the sign and magnitude of the scalar-invariant form factors. On the other hand, the contribution of scalar form factor $A_-(q^2)$ in $B_c\to \psi(2S,3S)/D^*(2S,3S)$ decays is constructive for which $\tilde{h}_t$ decreases when $A_-(q^2)$ is switched off; as shown in Fig.5.	
	\begin{figure}[!hbt]
		\includegraphics[width=0.4\textwidth]{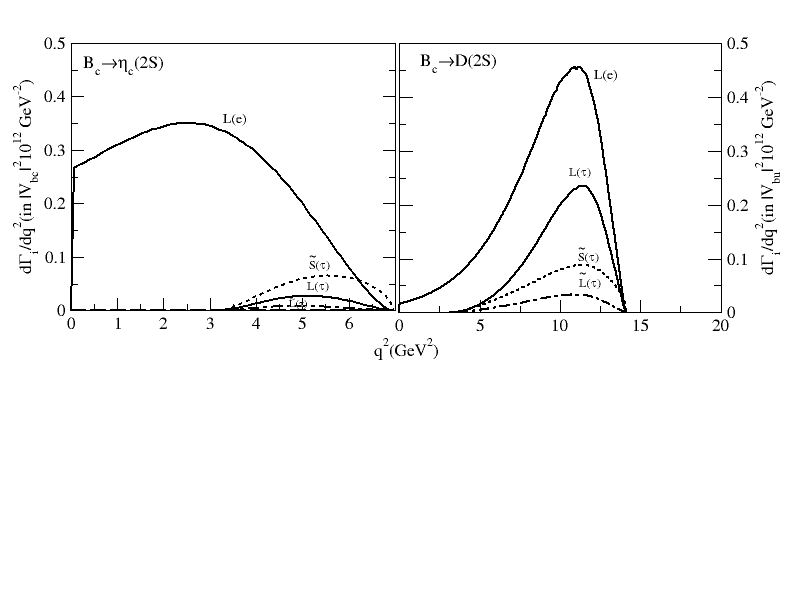}
		\includegraphics[width=0.4\textwidth]{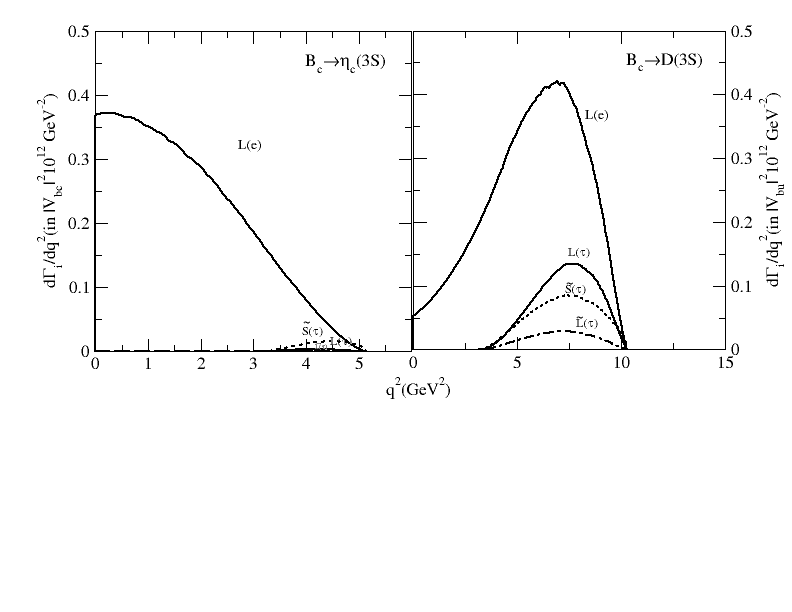}
		\caption{Partial helicity rates $\frac{d\Gamma_i}{dq^2}$ and $\frac{d\tilde{\Gamma_i}}{dq^2}$ as functions of $q^2$ for semileptonic decay rates for $B_c \to \eta_c(2S)/D(2S)$ and $B_c \to \eta_c(3S)/D(3S)$ decays.}
\end{figure}	

	In Fig.6 we plot the $q^2$-spectra for different helicity components in $B_c\to \eta_c(e^-,\tau^-)$ and $B_c\to D(e^-,\tau^-)$ decay modes. As in the case of semileptonic $B_c$ decays to ground state charmonium and charm mesons \cite{A71}, we also find here a considerable reduction of the longitudinal no-flip contribution in the $\tau^-$ modes compared to the $e^-$ modes. Besides contribution of the longitudinal no-flip part the scalar flip components over the allowed kinematic range provides a sizable contribution.
	\begin{figure}[!hbt]
		\resizebox{1\hsize}{!}{\includegraphics[width=0.4\textwidth]{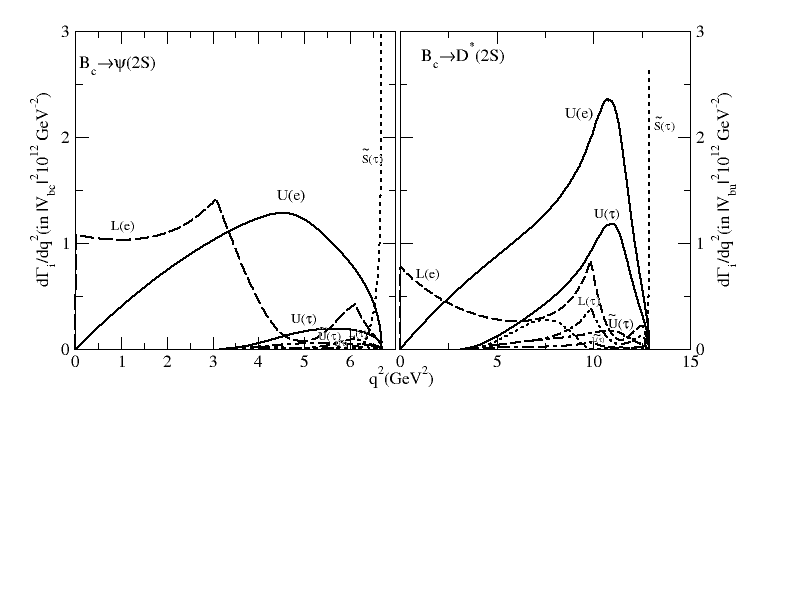}}
		\resizebox{1\hsize}{!}{\includegraphics[width=0.4\textwidth]{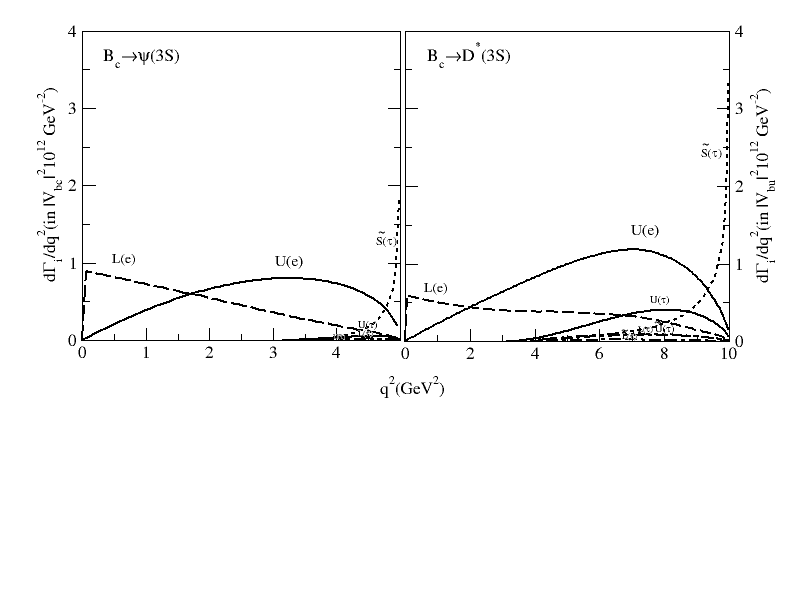}}
		\caption{Partial helicity rates $\frac{d\Gamma_i}{dq^2}$ and $\frac{d\tilde{\Gamma_i}}{dq^2}$ as functions of $q^2$ for semileptonic decay rates for $B_c \to \psi(2S)/D^*(2S)$ and $B_c \to \psi(3S)/D^*(3S)$ decays.}
	\end{figure}

Fig.7 depicts the $q^2$-spectra for different helicity components in $B_c\to \psi(e^-,\tau^-)$ and $B_c \to D^*(e^-,\tau^-)$ modes. We also find here that the spin-flip parts are negligible compared to no-flip parts as seen in our earlier study \cite{A71}. Besides contribution of longitudinal no-flip part, the dominant contribution also comes here from the transverse no-flip parts. The helicity rates are found reduced uniformly going from $e^-$ to $\tau^-$ modes, as expected. A significant contribution of the time like scalar-current via $\tilde{S}(\tau)$ is obtained in the high $q^2$-region.
	\begin{figure}[!hbt]
		\includegraphics[width=0.4\textwidth]{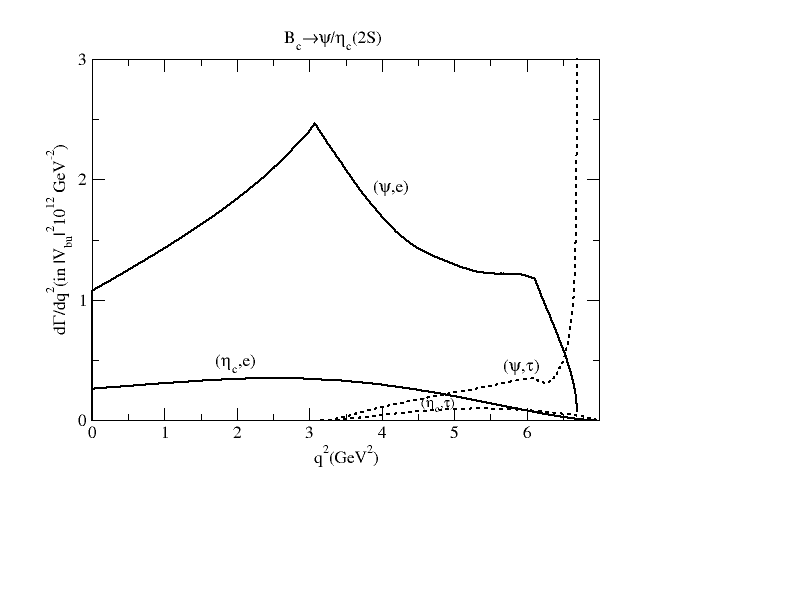}
		\includegraphics[width=0.4\textwidth]{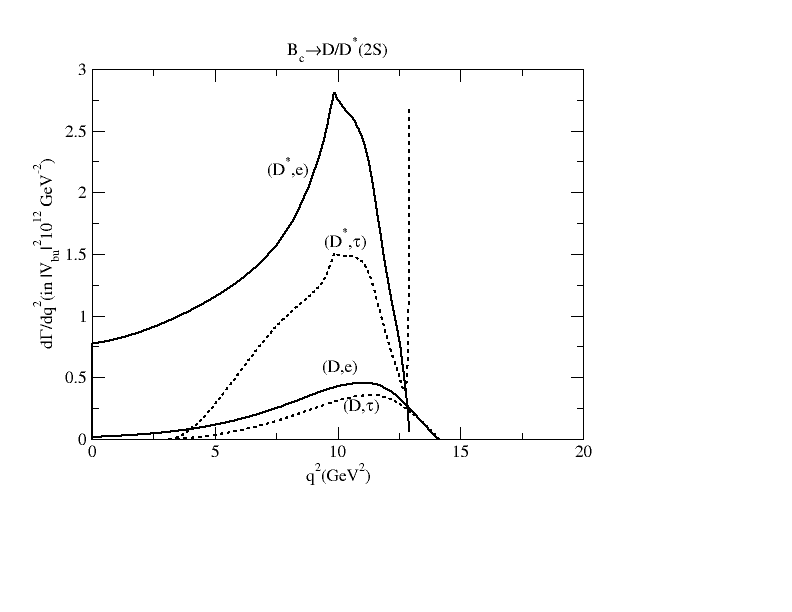}
		\caption{$q^2$-spectrum of semileptonic decay rates  $B_c \to \eta_c(2S)/\psi(2S)$ and $B_c \to D(2S)/D^*(2S)$ decays.}
	\end{figure}
	\begin{figure}[!hbt]
		\includegraphics[width=0.4\textwidth]{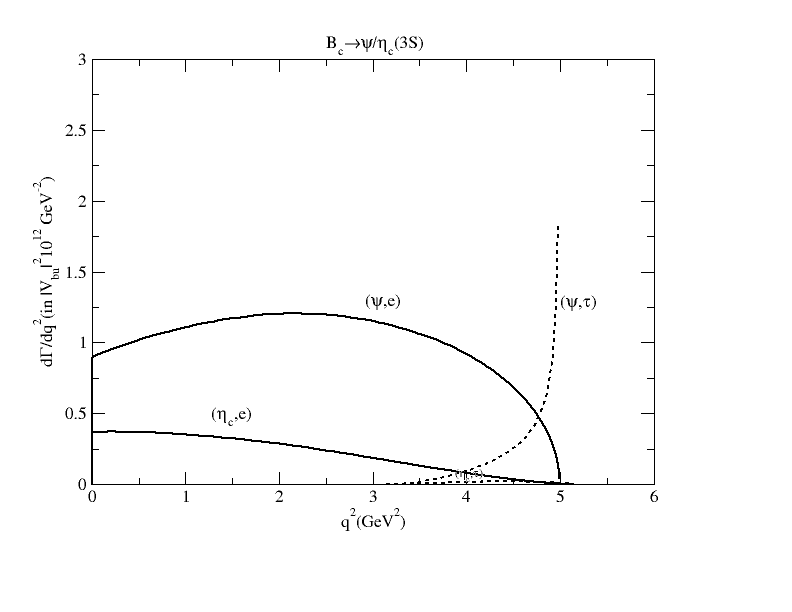}
		\includegraphics[width=0.4\textwidth]{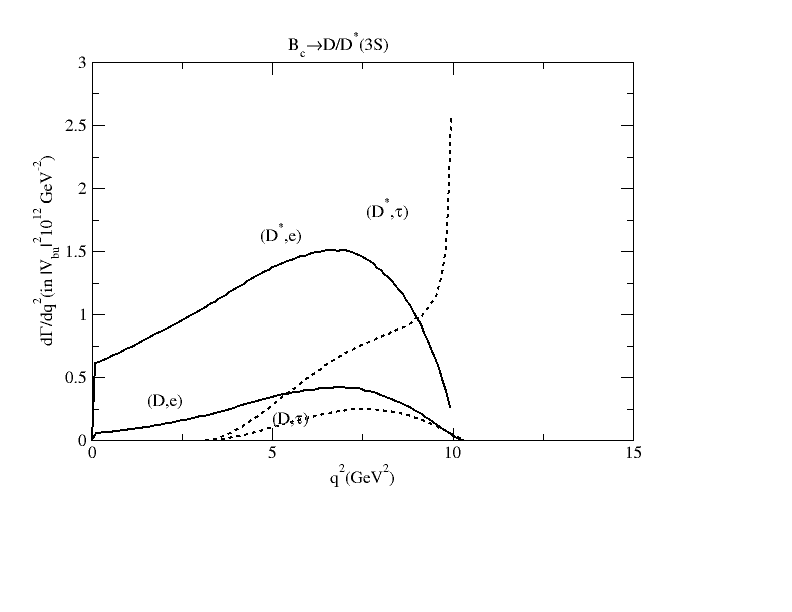}
		\caption{$q^2$-spectrum of semileptonic decay rates for $B_c \to \eta_c(3S)/\psi(3S)$ and $B_c \to D(3S)/D^*(3S)$ decays.}
	\end{figure}

	In Figs. 8 and 9 we plot the total $q^2$-spectra: $\frac{d\Gamma}{dq^2}$ for $B_c\to \eta_c(\psi)$ and $B_c\to D(D^*)$ decays, respectively, in their $e^-$ and $\tau^-$ modes. For $B_c\to \eta_c(2S)$ decay in $e^-$ mode, the $q^2$-spectra increases rapidly at $q^2\to0$ to a peak and then flattens with a slow rise to another peak near $q^2\simeq 2.5 \ GeV^2$. Thereafter it decreases slowly to zero value at $q^2\simeq 7\ GeV^2 $. In the $\tau^-$-mode, the spectra, which begin at a positive $q^2\simeq 3.3\ GeV^2$ away from $q^2\to 0$, increase slowly to a peak at $q^2\simeq 5.5\ GeV^2$ and then decrease almost at the same rate to zero at $q^2\simeq 7\ GeV^2$. The $\tau^-$-mode spectra in these decays is found almost below $e^-$-mode spectra throughout only to overcome slightly above the $e^-$-spectra towards the end of the kinematic range: $q^2\ge 6 \ GeV^2$.\\
	
	A similar trend of $q^2$-spectra is observed for $B_c\to \eta_c(3S)$ decays in their $e^-$ and $\tau ^-$ modes in the kinematic range of $0\leq q^2\le5\ GeV^2$ and $3.3\leq q^2\leq 5\ GeV^2$, respectively. For $B_c\to D(2S)$ decays, the $e^-$ mode spectra originating at $q^2\to 0$ rise slowly to a peak at about $q^2\simeq11.25\ GeV^2$ and then decrease to zero at $q^2\simeq13.3\ GeV^2$. On the other hand, the $\tau^-$-mode spectra for these decays starts at $q^2\simeq 3\ GeV^2$, slowly rise to a peak at $q^2\simeq 11.25\ GeV^2$, and fall to zero thereafter at $q^2\simeq13.8\ GeV^2$ lying all along within its $e^-$-mode spectra. For $B_c\to D(3S)$ decays the $e^-$ mode spectra rises fast near $q^2\simeq 0$, then rises moderately to a peak at $q^2\simeq 6.75\ GeV^2$ and thereafter falls to zero at $q^2\simeq 10.25\ GeV^2$. Its $\tau$-mode spectra originates at about $q^2\simeq 3\ GeV^2$ away from $q^2\to 0$ and then rises to the peak at $q^2\simeq7.5\ GeV^2$ after which it decreases to zero at $q^2=10.25\ GeV^2$, lying all along within its $e^-$-mode spectra.\\
	
	For $B_c\to \psi(2S)$ decays we find a sharp rise of $e^-$mode spectra to a peak at $q^2\to 0$ which rises thereafter to a prominent shoulder at $q^2\simeq3\ GeV^2$ beyond which it decreases in the region $3\le q^2\le 5.5\ GeV^2$. Then the spectra flatten up to $q^2\simeq 6\ GeV^2$ followed by its sharp fall to zero at about $6.7\ GeV^2$. Its $\tau ^-$-mode spectra originating at about $q^2\simeq 3.2\ GeV^2$ attains a slow linear rise up to $q^2\simeq 6\ GeV^2$; after which it slows down a bit only to hook up to a very high value at $q^2\simeq 6.7\ GeV^2$. On the other hand for $B_c\to \psi (3S)$ decays in $e^-$ mode, the spectra, which begins with a sharper rise to a peak at $q^2\to 0$, flattens to attain a peak at $q^2\simeq 2.3\ GeV^2$ and then decreases rapidly to zero at $q^2\simeq 5\ GeV^2$. In its $\tau^-$-mode, the spectra begins only at $q^2\simeq3.25\ GeV^2$ and increases slowly up to $q^2\simeq4.5\ GeV^2$ after which it attains a steep rise to a high value at about $q^2\simeq4.75\ GeV^2$.\\
		\begin{figure}[!hbt]
		\includegraphics[width=0.4\textwidth]{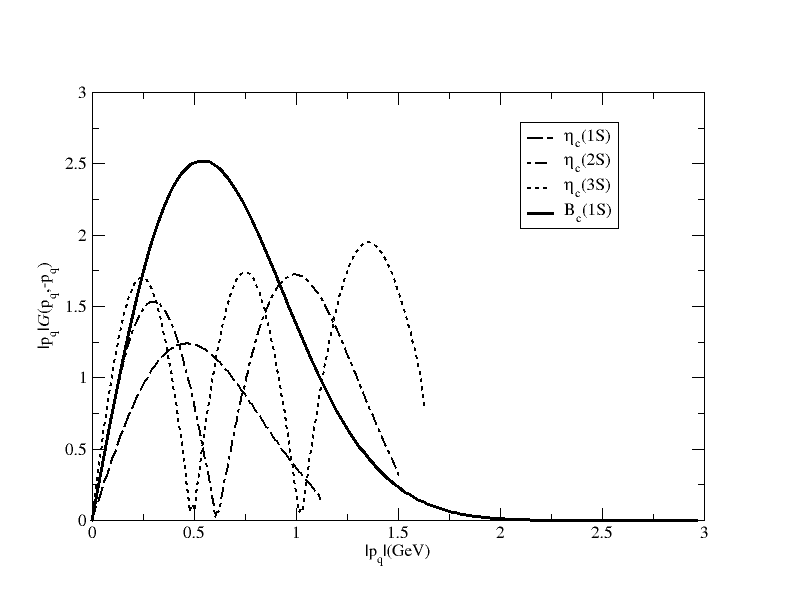}
		\includegraphics[width=0.4\textwidth]{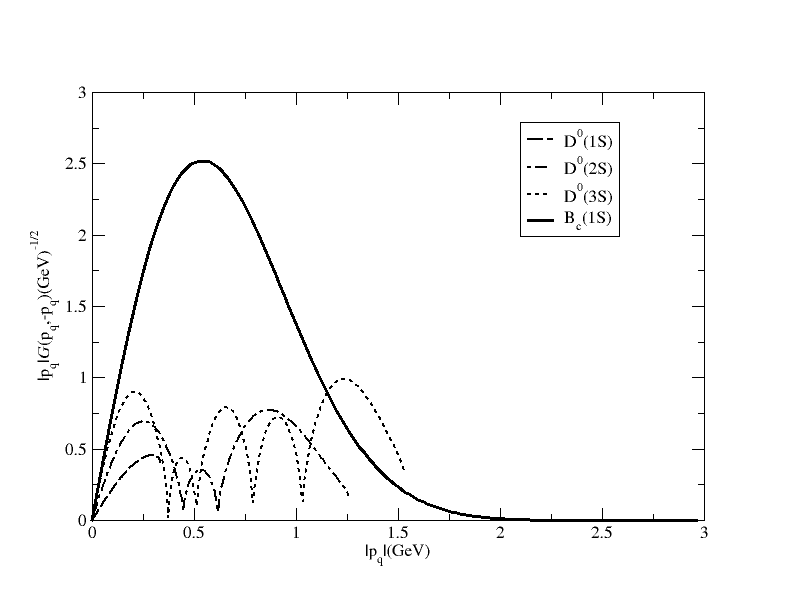}
		\includegraphics[width=0.4\textwidth]{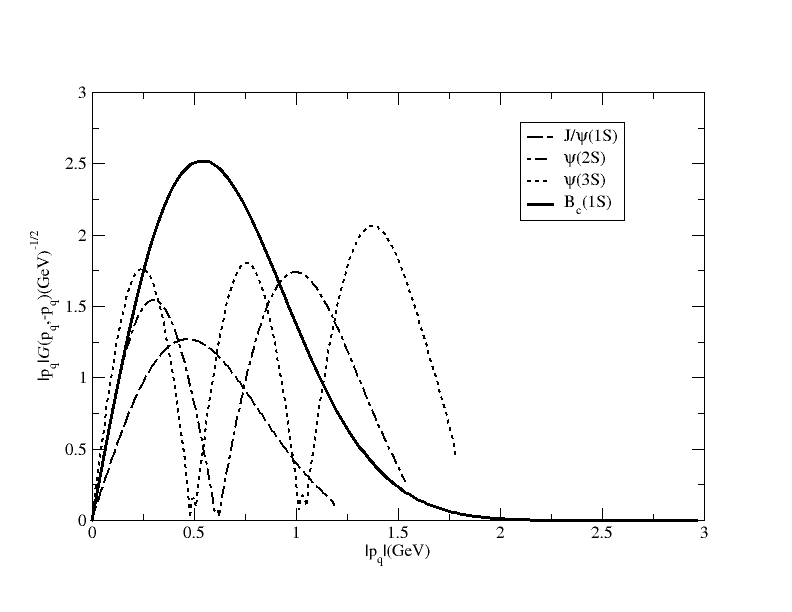}
		\includegraphics[width=0.4\textwidth]{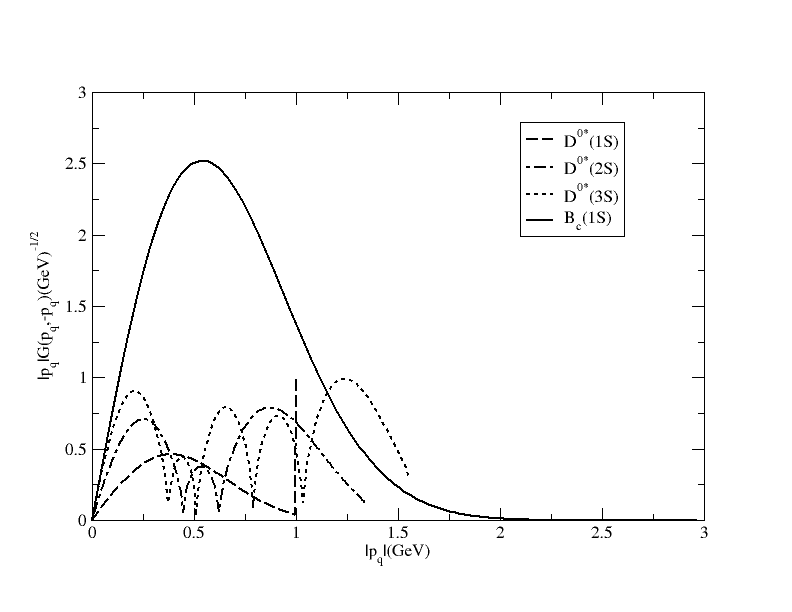}
		\caption{Overlap of momentum distribution amplitudes of the initial and final meson state.}
	\end{figure}
	
	For $B_c\to D^*(2S)$ decays, the $e^-$ mode spectra attains a step rise to a peak at $q^2\to 0$. Then the spectra increases moderately up to $q^2\simeq5\ GeV^2$ beyond which it increases rapidly to develop a sharp shoulder only to fall rapidly to a very small value near $q^2\simeq 13\ GeV^2$. Its $\tau^-$ mode spectra appears identical in shape through out the kinematic range lying all along within the $e^-$-spectra as expected except at $q^2\simeq13\ GeV^2$ where it attains a steep rise to a very high value at $q^2\simeq 13\ GeV^2$. For $B_c\to D^*(3S)$ decays the $e^-$-mode spectra rises swiftly at $q^2\to 0$ to a peak and then rises moderately till it attains a peak at about $q^2\simeq7\ GeV^2$. Thereafter it decreases to a very low value at $q^2\simeq10\ GeV^2$. Its $\tau^-$-mode spectra on the other hand begins also at about $q^2\simeq 3\ GeV^2$ away from $q^2\to 0$ and rises afterward till it overtakes its $e^-$-mode spectra at about $q^2\simeq 9\ GeV^2$. After overtaking the $e^-$-mode spectra, it attains a steep rise to a very high value. It may be noted here that in $B_c\to \psi (2S,3S)$ and $B_c\to D^*(2S,3S)$ decays, the shapes of the $\tau^-$-mode spectra are significantly reduced compared to that in the $e^-$-mode. However, towards the end of the kinematic region, the respective $\tau^-$-mode spectra are found to overtake the $e^-$-mode spectra contrary to the naive phase space expectations.
	
	Before evaluating the physical quantities of interest it is interesting to study the behavior of radial quark momentum distribution amplitude function of the decaying $B_c$ meson state together with those of the final $S$-wave charmonium and charm meson states. The behavior of momentum distribution amplitudes is shown in Fig.10. Note that the overlap regions between the momentum distribution amplitude function of the initial $B_c$ meson and final charmonium and charm meson states are obtained in the decreasing order from $1S$ to higher $2S$ and $3S$ final states. It is expected that contribution to decay rate should also be in decreasing order as on goes from ground to higher $2S$ and $3S$ final states.\\ 
	
	In Tables \ref{tab2} and \ref{tab3} we list our results for the integrated partial helicity rates: $\Gamma_i(i=U,L,P)$ and $\tilde{\Gamma}_i(i=U,L,S,SL)$ as well as the total decay rates of semileptonic $B_c$ decays. The partial tilde decay rates for each decay process in the $e^-$ mode are found tiny as expected from Eq.(12) and can therefore be neglected. But corresponding rates in $\tau^-$-mode, obtained comparable to the untilde rates, can hardly be neglected. From the contributions of individual helicity rates, we predict the decay rates for each process, in its $e^-$ as well as $\tau^-$ mode. Like all other model predictions, our predicted decay rates in $\tau^-$-modes are found, in general, smaller than that in $e^-$-modes. For $B_c\to \psi(2S)$ and $B_c\to \psi(3S)$ transitions, our predicted decay rate in $\tau^-$ mode is suppressed by a factor of $\sim 12$ and $\sim 13$, respectively. However, for $B_c\to \eta_c(2S)$ and $B_c\to \eta_c(3S)$ transitions the suppression is by a factor of $\sim 7$ and $\sim 43$, respectively compared to their respective decays in the $e^-$ mode. For the CKM suppressed transitions: $B_c\to D(2S)$ and $B_c\to D^*(2S)$, the $\tau^-$ modes are suppressed by a factor of $\sim 2$ whereas for $B_c\to D(3S)$ and $B_c\to D^*(3S)$ the suppression is by a factor of $\sim 2.5$ over corresponding $e^-$ modes.    
	\begin{table*}[!hbt]
		\renewcommand{\arraystretch}{1}
		\centering
		\setlength\tabcolsep{2.5pt}
			\setlength\tabcolsep{8.0pt}
		\caption{Helicity rates(in $10^{-15}$ GeV) of semileptonic $B_c$-meson decays into $2S$ charmonium and charm-meson states:}
		\label{tab2}
		\begin{tabular}{|c|c|c|c|c|c|c|c|c|c|}
			\hline Decay mode &$U$ & $\tilde{U}$ & $L$ & $\tilde{L}$ &$P$&$S$ & $\tilde{S}$&$\tilde{SL}$ & $\Gamma$ \\
			\hline${B^-_c}\to \eta_c(2S) e^-\nu_e$ & & & 2.877 & 5.275$10^{-7}$& & &17.57$10^{-7}$&5.51$10^{-7}$&2.877\\
			
			\hline$B_c\to\eta_c(2S) \tau^-\nu_\tau$ & & & 0.098 & 0.031 && &0.266 &0.0508&0.396\\
			
			\hline${B^-_c}\to \psi(2S) e^-\nu_e$ &9.236&4.430$\times10^{-7}$ &8.08&18.73$\times 10^{-7}$&5.685&3.166 &47.04$\times 10^{-7}$&16.63$\times 10^{-7}$&17.317\\
			
			\hline${B^-_c}\to \psi(2S) \tau^-\nu_\tau$ & 0.688 &0.206 &0.170& 0.050&0.392 &0.0309 &0.360&0.007&1.476\\
			
			\hline${B^-_c}\to D(2S) e^-\nu_e$ &  &  &0.042&8.86$\times 10^{-10}$& & &24.73$\times 10^{-10}$&8.51$\times 10^{-10}$&0.042\\
			
			\hline${B^-_c}\to D(2S)\tau^-\nu_\tau$ &  &&0.017&0.0027& &&0.0077&0.0026&0.027\\
			
			\hline${B^-_c}\to D^*(2S) e^-\nu_e$ &0.199&4.317$\times 10^{-9}$&0.070&10.17$\times 10^{-9}$&0.164&0.024&2.734$\times 10^{-8}$&7.187$\times 10^{-9}$&0.270\\
			
			\hline${B^-_c}\to D^*(2S)\tau^-\nu_\tau$ &0.069&0.011&0.014&0.0026&0.064&0.0093&0.0146&0.0028&0.113\\
			\hline
		\end{tabular}
	\end{table*}
	
	\begin{table*}[!hbt]
		\renewcommand{\arraystretch}{1}
		\centering
		\setlength\tabcolsep{2.5pt}
			\setlength\tabcolsep{8.0pt}
		\caption{Helicity rates(in $10^{-15}$ GeV) of semileptonic $B_c$-meson decays into $3S$ charmonium and charm-meson states:}
		\label{tab3}
		\begin{tabular}{|c|c|c|c|c|c|c|c|c|c|}
			\hline Decay mode &$U$ & $\tilde{U}$ & $L$ & $\tilde{L}$ &$P$&$S$ & $\tilde{S}$&$\tilde{SL}$ & $\Gamma$ \\
			\hline${B^-_c}\to \eta_c(3S) e^-\nu_e$ & & & 1.838 & 6.25$\times10^{-7}$& & &20.12$\times10^{-7}$&6.45$\times10^{-7}$&1.83\\
			
			\hline$B_c\to\eta_c(3S) \tau^-\nu_\tau$ & & & 0.007 & 0.002&& &0.032 &0.005&0.042\\
			
			\hline${B^-_c}\to \psi(3S) e^-\nu_e$ &4.678&3.09$\times10^{-7}$ &3.821&15.43$\times 10^{-7}$&1.767&0.347 &44.04$\times 10^{-7}$&13.442$\times 10^{-7}$&8.499\\
			
			\hline${B^-_c}\to \psi(3S) \tau^-\nu_\tau$ & 0.09 &0.032 &0.019& 0.0072&0.021 &0.188 &0.51&0.026&0.659\\
			
			\hline${B^-_c}\to D(3S) e^-\nu_e$ &  &  &0.036&16.17$\times 10^{-10}$& & &47.82$\times 10^{-10}$&16.03$\times 10^{-10}$&0.036\\
			
			\hline${B^-_c}\to D(3S)\tau^-\nu_\tau$ &  &&0.0075&0.0017& &&0.005&0.0017&0.014\\
			
			\hline${B^-_c}\to D^*(3S) e^-\nu_e$ &0.109&3.391$\times 10^{-9}$&0.05&8.091$\times 10^{-9}$&0.087&0.0206&2.326$\times 10^{-8}$&5.532$\times 10^{-9}$&0.160\\
			
			\hline${B^-_c}\to D^*(3S)\tau^-\nu_\tau$ &0.023&0.0051&0.0059&0.0014&0.0202&0.008&0.026&0.0024&0.0623\\
			\hline
		\end{tabular}
	\end{table*}

	From the predicted decay rates shown in Tables \ref{tab2} and \ref{tab3} and $B_c$-meson lifetime $\tau_{B_c}=0.510\ ps$, we predict the branching fractions (BF) for semileptonic $B_c\to \eta_c(\psi)$ and $B_c\to D(D^*)$ decays in their $e^-$ and $\tau^-$ modes. Our results for such decays to radially excited $2S$ and $3S$ final meson states are shown in Table \ref{tab4} and \ref{tab5}, respectively. Considering our predictions on semileptonic $B_c$ decays to charmonium and charm meson ground states \cite{A71}, we find that our predicted branching fractions for decays to $1S$, $2S$, and $3S$ final states are obtained in the hierarchy:
	\begin{equation}
		{\cal B}(B_c\to X(3S)l\nu_l)<{\cal B}(B_c\to X(2S)l\nu_l)<{\cal B}(B_c\to X(1S)l\nu_l)\nonumber
	\end{equation}
Our results for $B_c\to \eta_c(2S)$ and $B_c\to \eta_c(3S)$ decays in their $e^-$ mode are obtained $\sim 1.5$ and $\sim 2.5$ times down and in the $\tau ^-$ mode, the suppression is one and two orders of magnitude respectively, compared to our results in the final ground state charmonium. However, our predicted branching fractions for $B_c\to \psi(2S)$ and $B_c\to \psi(3S)$ decays in $e^-$ modes are $\sim2$ and $\sim4$ times down whereas, in $\tau$-mode, they are $1/2$ and $1$ order of magnitude down, respectively compared to those obtained for decays to ground state charmonium $(J/\psi)$.\\
 The node structure of the $2S$ wave function is responsible for small branching fractions. In the calculation of the overlap integral of meson wave functions, as there is no node for the initial wave functions, the contribution of positive and negative parts of the final wave function cancel each other out yielding a small branching fraction. About 3S final states, there are even more severe cancellations, which leads to still smaller branching fractions. As expected the tighter phase space and weaker $q^2$-dependence of form factors also lead to smaller branching fractions for transitions to higher excited $2S$ and $3S$ final states. Our predictions on branching fraction for decays to $2S$ final charmonium states in $e$ and $\tau $ modes are in reasonable agreement with those of Refs. \cite{A26,A55,A67} and about $1$ order of magnitude higher than the predictions in Refs.\cite{A39,A40,A41}. Although our predictions for decays to $3S$ charmonium states in their $\tau$-modes are found over-estimated compared to other model predictions shown in Table. IV, our predictions for corresponding decay in $e^-$-mode agree with those of \cite{A41, A55}. As in all other model predictions we find our prediction for $B_c\to \psi(2S)$ and $B_c\to \psi(3S)$ is the largest of all decay modes considered in the present study.\\
	\begin{table*}[!hbt]
		\centering
			\setlength\tabcolsep{8.0pt}
		\caption{Branching fractions(in\ \%) of semileptonic $B_c$ decays into $2S$ and $3S$ charmonium states:}
		\label{tab4}
		\begin{tabular}{|c|c|c|c|c|c|c|c|c|c|c|}
			\hline Decay mode&This work&\cite{A22}&\cite{A35}&\cite{A36}&\cite{A37}&\cite{A38,A39}&\cite{A48}&\cite{A51}&\cite{A63}&\cite{A80}\\
			\hline$B_c\to \eta_c(2S)e\nu$&0.22&0.056&0.046&0.07&0.0496&0.03&0.02&0.77&0.11&0.046\\
			$B_c\to \eta_c(2S)\tau \nu$&0.03&-&-&-&0.0025&-&-&0.053&0.0081&0.0013\\
			\hline $B_c\to \psi(2S) e\nu$&1.341&0.112&0.014&0.1&0.081&0.03&0.12&1.2&-&0.21\\
			$B_c\to \psi(2S) \tau\nu$&0.114&-&-&-&0.0408&-&-&0.084&-&0.015\\
			\hline$B_c\to \eta_c(3S)e\nu$&0.14&-&-&-&0.00414&5.5$\times 10^{-4}$&-&0.14&1.9$\times 10^{-2}$&-\\
			$B_c\to \eta_c(3S)\tau \nu$&0.003&-&-&-&0.0043$\times 10^{-2}$&5.0$\times 10^{-7}$&-&1.9$\times 10^{-4}$&5.7$\times 10^{-4}$&-\\
			\hline $B_c\to \psi(3S) e\nu$&0.658&-&-&-&0.0109&5.7$\times 10^{-4}$&-&3.6$\times 10^{-2}$&-&-\\
			$B_c\to \psi(3S) \tau\nu$&0.0511&-&-&-&0.010$\times 10^{-2}$&3.6$\times 10^{-6}$&-&3.8$\times 10^{-5}$&-&-\\
			\hline
		\end{tabular}
	\end{table*}
	
	Our results for $B_c\to D(2S)/D^*(2S)$ and $B_c\to D(3S)/D^*(3S)$ transitions are shown in Table \ref{tab5}. We find the branching fractions for decays in their $e^{-}$ as well as $\tau^- $ modes in the decreasing order as one goes from $1S$ to higher $2S$ and $3S$ final charm meson states as expected. For $B_c\to D$ decays in their $e^-$ as well as $\tau^-$ mode, the predicted branching fractions decrease marginally. However, for $B_c\to D^*$ decays in $e^-$ mode our results decreases marginally from $1S$ to higher $2S$ and $3S$ states, but in $\tau^-$ mode our predictions for $2S$ and $3S$ final states decrease $\sim 3$ and $\sim 5$ times, respectively, as compared to that for $1S$ final state.
	\begin{table*}[!hbt]
		\centering
			\setlength\tabcolsep{8.0pt}
		
		\caption{Branching fractions(in\ \%) of semileptonic $B_c$ decays into $2S$ and $3S$ charm meson states:}
		\label{tab5}
		\begin{tabular}{|c|cc|cc|cc|cc|}
			\hline Decay& \multicolumn{2}{c}{\ \ $B_c\to D(2S)$\ \ }\vline & \multicolumn{2}{c}{\ \ $B_c\to D^*(2S)$\ \ } \vline&\multicolumn{2}{c}{\ \ $B_c\to D(3S)$\ \ }\vline&\multicolumn{2}{c}{\ \ $B_c\to D^*(3S)$\ \ }\vline\\
			\hline
			Mode&$e^-$&$\tau^-$&$e^-$&$\tau^-$ &$e^-$&$\tau^-$&$e^-$&$\tau^-$\\
			\hline This work&0.0032&0.0021&0.020&0.0087&0.0028&0.0013&0.012&0.0048\\
			\hline
		\end{tabular}
	\end{table*}

\begin{table}[hbt!]
	\centering
	\caption{Forward-backward asymmetry $A_{FB}$ and the  asymmetry parameter $\alpha^*$ for semileptonic $B_c$-decays to $2S$ charmonium and charmed meson states.}
	\label{tab6}
	\begin{tabular}{|c|c|c|c|}
		\hline Decay process &$A_{FB}(l^-)$&$A_{FB}(l^+)$&$\alpha^*$\\
		\hline$B_c\to \eta_c(2S)e\nu$&5.75$\times10^{-7}$&5.75$\times10^{-7}$&\\
		$B_c\to \eta_c(2S)\tau \nu$&0.384&0.384&\\
		\hline $B_c\to \psi(2S) e\nu$&0.246&-0.246&-0.272\\
		$B_c\to \psi(2S) \tau\nu$&0.183&-0.214&-0.130\\
		\hline$B_c\to D(2S)e\nu$&6.051$\times10^{-8}$&6.051$\times10^{-8}$&-\\
		$B_c\to D(2S)\tau \nu$&0.287&0.287&\\
		\hline $B_c\to D^*(2S) e\nu$&0.457&0.457&0.172\\
		$B_c\to D^*(2S)\tau\nu$&0.348&-0.5003&0.116\\
		\hline
	\end{tabular}
\end{table}
\begin{table}[hbt!]
	\centering
	\caption{Forward-backward asymmetry $A_{FB}$ and the  asymmetry parameter $\alpha^*$ for semileptonic $B_c$-decays to $3S$ charmonium and charmed meson states.}
	\label{tab7}
	\begin{tabular}{|c|c|c|c|}
		\hline Decay process &$A_{FB}(l^-)$&$A_{FB}(l^+)$&$\alpha^*$\\
		\hline$B_c\to \eta_c(3S)e\nu$&1.05$\times10^{-6}$&1.05$\times10^{-6}$&\\
		$B_c\to \eta_c(3S)\tau \nu$&0.367&0.367&\\
		\hline $B_c\to \psi(3S) e\nu$&0.155&-0.155&-0.240\\
		$B_c\to \psi(3S) \tau\nu$&-0.095&-0.144&-0.794\\
		\hline$B_c\to D(3S)e\nu$&1.329$\times10^{-7}$&1.329$\times10^{-7}$&-\\
		$B_c\to D(3S)\tau \nu$&0.355&0.355&\\
		\hline $B_c\to D^*(3S) e\nu$&0.4064&-0.4064&0.038\\
		$B_c\to D^*(3S)\tau\nu$&0.123&-0.362&-0.404\\
		\hline
	\end{tabular}
\end{table}
Using appropriate helicity rates for different decay modes, we evaluate two quantities of interest: the forward-backward asymmetry $A_{FB}$ $(14)$ and asymmetry parameter $\alpha^*$ $(15)$. Our predicted $A_{FB}$ and $\alpha^*$ for decays to $2S$ and $3S$ final meson states are shown in Table \ref{tab6} and \ref{tab7}, respectively. For decays to spin 0 state, $A_{FB}$, which is proportional to the helicity amplitude $\tilde{SL}$, is found tiny in $e^-$ mode but non-negligible in $\tau^-$ mode. Looking at the expression for $A_{FB}\ (14)$ it is easy to relate corresponding parameters in the $e^-$ and $e^+$ mode in the form: $A_{FB}(e^-)=-A_{FB}(e^+)$ and $A_{FB}(\tau^-)\ne-A_{FB}(\tau^+)$ for decays to spin 1 state. To determine the transverse and longitudinal components of the final state vector mesons in $B_c\to \psi$ and $B_c\to D^*$ transitions in their $e^-$ and $\tau^-$ modes, we calculate the asymmetry parameter $\alpha^*$. One can see from Tables \ref{tab2} and \ref{tab3} that the transverse and longitudinal pieces in $B_c\to \psi(2S)$ and $B_c\to \psi(3S)$ decays are found almost equal in $e^-$ mode. But the transverse component dominates over longitudinal parts in their $\tau^-$ mode by a factor of $\sim 4$. In the $B_c\to D^*(2S)$ and $B_c\to D^*(3S)$ decays, the transverse components dominate over longitudinal parts by a factor $\sim 2$ in $e^-$ mode. However, in the $\tau^-$ mode, transverse components dominate over longitudinal parts by a factor of $\sim 5$ and $\sim 4$ for $B_c\to D^*(2S)$ and $B_c\to D^*(3S)$, respectively. The asymmetry parameter is found close to $-27\%$ in $e^-$ mode and $-13\%$ in $\tau ^-$ mode for $B_c\to \psi(2S)$ decays. For $B_c\to D^*(2S)$ decays it is found close to $17\%$ and $12\%$ in $e^-$ and $\tau^-$ mode, respectively. In $B_c\to \psi(3S)$ decays, $\alpha^*$ is obtained close to $-24\%$ and $-79\%$ in their $e^-$ and $\tau^-$ mode, respectively, whereas in $B_c\to D^*(3S)$ decays, it is found to be $\sim 4\%$ in $e^-$ mode and $-40\%$ in $\tau^-$ mode. This is because the transverse components of helicity amplitudes provide significant contributions compared to the tiny contributions of the longitudinal and scalar parts for decays in $e^-$ mode. On the other hand, the scalar flip part $\tilde{S}$ of the helicity amplitude dominates over the rest part and contributes destructively in the decays in their $\tau^-$ modes yielding a highly suppressed asymmetry parameter $\alpha^*$ as low as $-79\%$ and $-40\%$ in $B_c\to \psi(3S)$ and $B_c\to D^*(3S)$ decays, respectively.\\

\begin{table*}[!hbt]
	\centering
	\caption{Ratios of branching fractions for semileptonic $B_c$-decays to radially excited charmonium states}
	\label{tab8}
	\begin{tabular}{|c|c|c|c|c|c|c|}
		\hline Ratio&This work&\cite{A37}&\cite{A38,A39}&\cite{A59}&\cite{A63}&\cite{A80}\\
		\hline${\cal R}_{\eta_c}(2S)$&7.33&18.4&-&14.5&1.35&35.38\\
		\hline${\cal R}_{\eta_c}(3S)$&46.67&96.24&1.1$\times 10^3$&7.36$\times 10^2$&33.33&\\
		\hline${\cal R}_{\psi}(2S)$&11.76&1.98&-&14.3&-&14\\
		\hline${\cal R}_{\psi}(3S)$&12.876&109&158.33&947.4&&\\
		\hline
	\end{tabular}
\end{table*}

Finally, we evaluate the observable: ${\cal R}=\frac{{\cal B}(B_c\to X(nS)l\nu_l)}{{\cal B}(B_c\to X(nS)\tau\nu_{\tau})}$ and our results are listed in Table \ref{tab8} in comparison with other model predictions. The CKM matrix elements do not contribute to the ratio ${\cal R}$. The uncertainties due to the model calculation and CKM parameter etc. are cancelled in the evaluation of the observable ${\cal R}$. Therefore, contrary to other observables, the ratio ${\cal R}$, in which the production of $B_c$ meson is cancelled totally, provides an essential test of the phenomenological model used in the description of decay processes. Our predicted ${\cal R}$ for $B_c$ decays to charmonium states are comparable to other standard model predictions as shown in Table \ref{tab8}. However, in the absence of predicted data from established model approaches in the literature for $B_c$ decays to charm meson states, our predictions: ${\cal R}_{D}(2S)=1.523,\ {\cal R}_{D}(3S)=2.153,\ {\cal R}_{D^*}(2S)=2.298$ and ${\cal R}_{D^*}(3S)=2.5$ can be useful to identify the $B_c$-channels characterized by clear experimental signature. The departure of the SM predictions on ${\cal R}$ from the observed data would highlight the puzzle in flavor physics and violation of lepton flavor universality(LFU) hinting at the new physics beyond SM.

\section{Summary and conclusion}
In this paper, we study the exclusive $B_c$-meson decays into radially excited $2S$ and  $3S$ charmonium and charm meson states in the framework of the RIQ model based on an average flavor-independent confining potential in an equally mixed scalar-vector harmonic form. The invariant form factors representing decay amplitudes are extracted from the overlapping integral of meson wave functions derivable from the model dynamics.\\
 Since our main objective here is to evaluate the lepton mass effects on the semileptonic $B_c$-meson decays, we analyze $B_c\to \eta_c/\psi (nS)l\nu $ and $B_c\to D/D^*(nS)l\nu$ in their $e^-$ and $\tau^-$ modes separately. For this, we study the $q^2$-dependence of relevant physical quantities such as the invariant form factors, helicity amplitudes, partial helicity rates, and the total $q^2$-spectra for all decay processes analyzed in the present work. Before evaluating the decay rates we study the behavior of radial quark momentum distribution amplitude function of the decaying $B_c$ meson state together with those of the final $S$-wave charmonium and charm meson states. From the overlapping regions, one could assess the decay rates of different $B_c$-meson decay modes involving daughter mesons from ground(1S) to 2S and 3S higher excited states.

Considering contribution from different partial helicity rates: $\frac{d\Gamma_i}{dq^2}\ (i=U,L,P)$ and $\frac{d\tilde{\Gamma}_i}{dq^2}\ (i=U,L,S,SL)$, the total $q^2$-spectrum $\frac{d\Gamma}{dq^2}$ is obtained for each decay process, from which we predict the decay rates/branching fractions of the semileptonic $B_c$-meson decays to their $e^-$ as well as $\tau^-$-modes. Our predictions on the decay rates/ branching fractions for decay processes are found in overall agreement with other SM predictions. The decay rate for $B_c\to\psi(2S) $ as well as $B_c\to \psi(3S)$ transition is found largest of all. As expected, the decay rates for $B_c\to \eta_c/\psi(nS)$ decay induced by $b\to c$ transition at the quark level dominate over those for $B_c\to D/D^*(nS)$ decays induced by the quark level $b\to u$ transition in their respective $e^-$ as well as $\tau^-$ modes. The contributions of longitudinal and transverse flip parts in the $\tau$ mode dominate over $e^-$ mode contribution as shown in Table 2. However, considering the contribution from no-flip parts, the total contribution to decay rate/branching fractions for all decays in the $e^-$ mode dominates over corresponding $\tau$ mode contribution. Like all other model predictions our predicted branching fractions for semileptonic $B_c$ decays to charmonium and charm meson ground \cite{A71} as well as $ 2S$ and $3S$ final states are obtained in the hierarchy:
\begin{equation}
	{\cal B}(B_c\to X(3S)l\nu_l)<{\cal B}(B_c\to X(2S)l\nu_l)<{\cal B}(B_c\to X(1S)l\nu_l)\nonumber
\end{equation}
We also find that the $\tau^-$ modes for all decays are suppressed in comparison with corresponding $e^-$ modes.

Using appropriate helicity rates we evaluate two quantities of interest:(1)Forward-backward asymmetry ‘$A_{FB}$’ and the asymmetry parameter ‘$\alpha^*$’. $A_{FB}$ in the present analysis is found negligible for transitions into spin 0 state in their $e^-$-mode but non-negligible in the $\tau^-$-mode, as expected. For transition into spin 1 state, we also find $A_{FB}(e^-)=-A_{FB}(e^+)$ and $A_{FB}(\tau^-)\ne -A_{FB}(\tau ^+)$. We evaluate the asymmetry parameter $\alpha^*$, which determines the transverse and longitudinal components of the final vector meson state in $B_c\to\psi(nS)$ and $B_c\to D^*(nS)$ transitions. We find the asymmetry parameter close to $-27\%$ in $e^-$ mode and $-13\%$ in $\tau ^-$ mode for $B_c\to \psi(2S)$ decays. For $B_c\to D^*(2S)$ decays it is found close to $17\%$ and $12\%$ in $e^-$ and $\tau^-$ mode, respectively. However, for $B_c\to \psi(3S)$ decays, $\alpha^*$ is obtained close to $-24\%$ and $-79\%$ in their $e^-$ and $\tau^-$ mode, respectively, whereas in $B_c\to D^*(3S)$ decays, it is found to be $\sim 4\%$ in $e^-$ mode and $-40\%$ in $\tau^-$ mode. This is because the transverse components of helicity amplitudes provide significant contribution compared to the tiny contributions of the longitudinal and scalar parts for decays in $e^-$ mode. On the other hand, the scalar flip part $\tilde{S}$ of the helicity amplitude dominates over the rest part and contributes destructively in the decays in their $\tau^-$ modes yielding a highly suppressed asymmetry parameter $\alpha^*$ as low as $-79\%$ and $-40\%$ in $B_c\to \psi(3S)$ and $B_c\to D^*(3S)$ decays, respectively.

Finally, we evaluate the observable ${\cal R}$, which corresponds to the ratios of branching fractions for transitions in $e^-$ modes to the corresponding value in $\tau^-$ modes. Our predicted observable ${\cal R}$ for $B_c\to \eta_c(nS)$ and $B_c\to \psi (nS)$ is found comparable to other SM predictions. The departure of the SM predictions of ${\cal R}$ from the experimental data highlights the puzzle in flavor physics and the failure of lepton flavor universality hinting at new physics beyond SM. However, in the absence of predicted data from established model approaches in the literature for $B_c$ decays to charm meson states, our predictions: ${\cal R}_{D}(2S),\ {\cal R}_{D}(3S),\ {\cal R}_{D^*}(2S)$ and ${\cal R}_{D^*}(3S)$ can be useful to identify the $B_c$-channels characterized by clear experimental signature. Given the possibility of high statistics $B_c$-events expected to yield up to $10^{10}$ events per year at the Tevatron and LHC, semileptonic $B_c$-meson decays into charmonium and charm mesons in their ground as well as excited states offer a fascinating area for future research.\\

 \begin{acknowledgements} 
	The library and computational facilities provided by the authorities of Siksha 'O' Anusandhan Deemed to be University, Bhubaneswar, 751030, India are duly acknowledged.
\end{acknowledgements}
\appendix
\section{CONSTITUENT QUARK ORBITALS AND MOMENTUM PROBABILITY AMPLITUDES}\label{app}

In the RIQ model, a meson is picturized as a color-singlet assembly of a quark and an antiquark independently confined by an effective and average flavor independent potential in the form:
$U(r)=\frac{1}{2}(1+\gamma^0)(ar^2+V_0)$ where ($a$, $V_0$) are the potential parameters. It is believed that the zeroth-order quark dynamics  generated by the phenomenological confining potential $U(r)$ taken in equally mixed scalar-vector harmonic form can provide an adequate tree-level description of the decay process being analyzed in this work. With the interaction potential $U(r)$ put into the zeroth-order quark lagrangian density, the ensuing Dirac equation admits a static solution of positive and negative energy as: 
\begin{eqnarray}
	\psi^{(+)}_{\xi}(\vec r)\;&=&\;\left(
	\begin{array}{c}
		\frac{ig_{\xi}(r)}{r} \\
		\frac{{\vec \sigma}.{\hat r}f_{\xi}(r)}{r}
	\end{array}\;\right)U_{\xi}(\hat r)
	\nonumber\\
	\psi^{(-)}_{\xi}(\vec r)\;&=&\;\left(
	\begin{array}{c}
		\frac{i({\vec \sigma}.{\hat r})f_{\xi}(r)}{r}\\
		\frac{g_{\xi}(r)}{r}
	\end{array}\;\right){\tilde U}_{\xi}(\hat r)
\end{eqnarray}
where, $\xi=(nlj)$ represents a set of Dirac quantum numbers specifying 
the eigen-modes;
$U_{\xi}(\hat r)$ and ${\tilde U}_{\xi}(\hat r)$
are the spin angular parts given by,
\begin{eqnarray}
	U_{ljm}(\hat r) &=&\sum_{m_l,m_s}<lm_l\;{1\over{2}}m_s|
	jm>Y_l^{m_l}(\hat r)\chi^{m_s}_{\frac{1}{2}}\nonumber\\
	{\tilde U}_{ljm}(\hat r)&=&(-1)^{j+m-l}U_{lj-m}(\hat r)
\end{eqnarray}
With the quark binding energy $E_q$ and quark mass $m_q$
written in the form $E_q^{\prime}=(E_q-V_0/2)$,
$m_q^{\prime}=(m_q+V_0/2)$ and $\omega_q=E_q^{\prime}+m_q^{\prime}$, one 
can obtain solutions to the resulting radial equation for 
$g_{\xi}(r)$ and $f_{\xi}(r)$in the form:
\begin{eqnarray}
	g_{nl}&=& N_{nl} (\frac{r}{r_{nl}})^{l+l}\exp (-r^2/2r^2_{nl})
	L_{n-1}^{l+1/2}(r^2/r^2_{nl})\nonumber\\
	f_{nl}&=& N_{nl} (\frac{r}{r_{nl}})^{l}\exp (-r^2/2r^2_{nl})\nonumber\\
	&\times &\left[(n+l-\frac{1}{2})L_{n-1}^{l-1/2}(r^2/r^2_{nl})
	+nL_n^{l-1/2}(r^2/r^2_{nl})\right ]
\end{eqnarray}
where, $r_{nl}= a\omega_{q}^{-1/4}$ is a state independent length parameter, $N_{nl}$
is an overall normalization constant given by
\begin{equation}
	N^2_{nl}=\frac{4\Gamma(n)}{\Gamma(n+l+1/2)}\frac{(\omega_{nl}/r_{nl})}
	{(3E_q^{\prime}+m_q^{\prime})}
\end{equation}
and
$L_{n-1}^{l+1/2}(r^2/r_{nl}^2)$ etc. are associated Laguerre polynomials. The radial solutions yields an independent quark bound-state condition in the form of a cubic equation:
\begin{equation}
	\sqrt{(\omega_q/a)} (E_q^{\prime}-m_q^{\prime})=(4n+2l-1)
\end{equation}
The solution of the cubic equation provides the zeroth-order binding energies of 
the confined quark and antiquark for all possible eigenmodes.

In the relativistic independent particle picture of this model, the constituent quark 
and antiquark are thought to move independently inside the $B_c$-meson bound state 
with momentum $\vec p_b$ and $\vec p_c$, respectively. Their momentum probability 
amplitudes are obtained in this model via momentum projection of respective quark orbitals (A1) in the following forms:

For ground state mesons:($n=1$,$l=0$)
\begin{eqnarray}
	G_b(\vec p_b)&&={{i\pi {\cal N}_b}\over {2\alpha _b\omega _b}}
	\sqrt {{(E_{p_b}+m_b)}\over {E_{p_b}}}(E_{p_b}+E_b)\nonumber\\
	&&\times\exp {(-{
			{\vec {p_b}}^2\over {4\alpha_b}})}\nonumber\\
	{\tilde G}_c(\vec p_c)&&=-{{i\pi {\cal N}_c}\over {2\alpha _c\omega _c}}
	\sqrt {{(E_{p_c}+m_c)}\over {E_{p_c}}}(E_{p_c}+E_c)\nonumber\\
	&&\times\exp {(-{
			{\vec {p_c}}^2\over {4\alpha_c}})}
\end{eqnarray}

\noindent In terms of individual momentum probability amplitudes: $G_b(\vec{p_b})$ and $G_c(\vec{p_c})$ of the constituent quarks, ${\cal G}_{B_c}(\vec{p_b},\vec{p}-\vec{p_b})$, the effective momentum distribution function is taken in the form:
\begin{equation}
	{\cal G}_{B_c}(\vec{p_b},\vec{p}-\vec{p_b})=\sqrt{G_b(\vec{p_b})G_c(\vec{p}-\vec{p_b})}
\end{equation}
in the straightforward extension of the ansatz of Margolis and Mendel in their bag model description \cite{A81}.

In the RIQ model, the wavepacket representing a meson bound state $\vert B_c(\vec{p}, S_{B_c})\rangle$, for example, at a definite momentum $\vec{p}$, and spin $S_{B_c}$ is taken in the form\cite{A67,A68,A69,A70,A71,A72}
\begin{equation}
	\vert B_c(\vec{p},S_{B_c})\rangle=\hat{\Lambda}(\vec{p},S_{B_c})\vert (\vec{p_b},\lambda_b);(\vec{p_c},\lambda_c)\rangle 
\end{equation}
where $\vert (\vec{p_b},\lambda_b);(\vec{p_c},\lambda_c)\rangle $ is the Fock space representation of the unbound quark and antiquark in a color-singlet configuration with their respective momentum and spin: $(\vec{p_b},\lambda_b)$ and $(\vec{p_c},\lambda_c)$. Here $\hat{b_b}^\dagger(\vec{p_b},\lambda_b)$  and $\hat{b_c}^\dagger(\vec{p_c},\lambda_c)$ denote the quark-antiquark creation operator, respectively, and $\hat{\Lambda}(\vec{p}, S_{B_c})$ is a bag like integral operator in the form:
\begin{eqnarray}
	\hat{\Lambda}(\vec{p},S_{B_c})=&&\frac{\sqrt{3}}{\sqrt{N(\vec{p})}}\sum_{\delta_b \delta_c}\zeta_{b_c}^{B_c}(\lambda_b,\lambda_c)\nonumber\\
	&&\int d^3p_bd^3p_c\delta^{(3)}(\vec{p_b}+\vec{p_c}+\vec{p}){\cal G}_{B_c}(\vec{p_b},\vec{p_c})
\end{eqnarray}
Here $\sqrt{3}$ is the effective color factor, $\zeta_{b_c}^{B_c}(\lambda_b,\lambda_c)$ is the $SU(6)$ spin-flavor coefficients for $B_c$-meson and $N(\vec{p})$ is the meson-state normalization obtained in an integral form:
\begin{equation}
	N(\vec{p})=\int d^3\vec{p_b}\vert {\cal G}_{B_c}(\vec{p_b},\vec{p}-\vec{p_b})\vert^2
\end{equation}
by imposing the normalization condition $\langle{B_c}(\vec{p})\vert{B_c}(\vec{p}^{'})\rangle=\delta^3 (\vec{p}-\vec{p}^{'})$.

The binding energy of constituent quark and antiquark in the parent and daughter meson in their ground as well as radially excited states are obtained by solving respective cubic equations representing appropriate bound state conditions(A5).

\end{document}